\documentstyle[11pt]{article}
\newlength{\abstractwidth}
\renewcommand{\thefootnote}{\fnsymbol{footnote}}
\renewcommand{\thanks}[1]{\footnote{#1}} 
\newcommand{\starttext}{
\setcounter{footnote}{0}
\renewcommand{\thefootnote}{\arabic{footnote}}}
\newcommand{\be}{\begin{equation}}
\newcommand{\bea}{\begin{eqnarray}}
\newcommand{\eea}{\end{eqnarray}}
\newcommand{\beq}{\begin{equation}}
\newcommand{\ee}{\end{equation}}
\newcommand{\eeq}{\end{equation}}
\newcommand{\N}{{\cal N}}
\newcommand{\<}{\langle}
\renewcommand{\>}{\rangle}
\def\ba{\begin{eqnarray}}
\def\ea{\end{eqnarray}}
\newcommand{\PSbox}[3]{\mbox{\rule{0in}{#3}\includegraphics{#1}\hspace{#2}}}

\begin{document}
\begin{titlepage}
\bigskip
\hskip 3.7in\vbox{\baselineskip12pt
\hbox{MIT-CTP-2727}\hbox{hep-th/9804058}}
\bigskip\bigskip\bigskip\bigskip

\centerline{\LARGE \bf Correlation functions in the ${\rm CFT}_{d}/AdS_{d+1}$
correspondence}

\bigskip\bigskip
\bigskip\bigskip

\centerline{ Daniel Z. Freedman$^{a,b,}$\footnote[1]{\tt dzf@math.mit.edu,}, 
Samir D. Mathur$^{a,}$\footnote[2]{\tt me@ctpdown.mit.edu,}, }
\medskip
\centerline{Alec Matusis$^{a,}$\footnote[3]{\tt alec\_m@ctp.mit.edu,}  
and Leonardo Rastelli$^{a,}$\footnote[4]{\tt rastelli@ctp.mit.edu.}}
\bigskip
\bigskip
\centerline{$^a$ \it Center for Theoretical Physics}
\centerline{ \it Massachusetts Institute of Technology}
\centerline{ \it Cambridge, {\rm MA}  02139}
\bigskip
\centerline{$^b$ \it Department of Mathematics}
\centerline{ \it Massachusetts Institute of Technology}
\centerline{\it Cambridge, {\rm MA} 02139}
\bigskip\bigskip

\begin{abstract}
Conformal techniques are applied to the calculation of integrals on
$AdS_{d+1}$ space which define correlators of composite operators in the 
superconformal field theory on the $d$--dimensional boundary. The 3--point
amplitudes for   scalar fields of arbitrary mass and  gauge fields
in the $AdS$ supergravity are calculated explicitly. For 3 gauge fields
we compare in detail with the known conformal structure of the $SU(4)$ flavor
current correlator $\< J_{i}^a J_{j}^b J_{k}^c \>$
of the $\N=4$, $d=4$  $SU(N)$ SYM theory. Results agree with the free
field approximation as would be expected 
from superconformal non--renormalization
theorems. In studying the Ward identity relating $\< J_i^a {\cal O}^I 
{\cal O}^J \>$ to $\< {\cal O}^I {\cal O}^J \> $ 
for (non--marginal) scalar composite operators ${\cal O}^I$,
we find that there is a subtlety in obtaining
the normalization of  $\< {\cal O}^I {\cal O}^J \> $ from the
supergravity action integral.
\baselineskip=16pt

\end{abstract}

\end{titlepage}
\starttext
\baselineskip=18pt
\setcounter{footnote}{0}
\section{Introduction}
The fact that the near horizon 
geometry~\cite{nearhorizongeometry}--\cite{nearhorizongeometryfinal}
 of typical brane configurations
in string/M theory is the product space $AdS_{d+1} \times S_p$ with
$d+1+p=10/11$ 
 has suggested an intriguing conjecture~\cite{maldacena}
 relating gauged supergravity
theory on $AdS_{d+1}$ with a superconformal theory on its $d$--dimensional
boundary~\cite{maldacena}. See 
also~\cite{earlierappearance}--\cite{earlierappearancefinal} 
for earlier appearance
of this correspondence in the context of  black hole physics and
\cite{everybodyfirst}--\cite{everybodylast} 
for recent relevant work on the subject.

 Precise forms of the conjecture~\cite{maldacena} have been stated and
investigated in~\cite{polyakov,witten} (see also~\cite{ferrarafronsdal})
for the $AdS_5 \times S_5$ geometry
of $N$ 3--branes in Type--IIB string theory. The superconformal theory
on the world--volume of the $N$ branes is  $\N=4$
SUSY Yang--Mills with gauge group $SU(N)$.
The conjecture holds in the limit 
of a large number 
$N$ of branes with $g_{st} N \sim g_{YM}^2 N$ fixed but large. As
$N \rightarrow \infty$ the string theory becomes weakly coupled and one can 
neglect string loop corrections; $N g_{st}$ large ensures that the $AdS$
 curvature is small so one can trust the supergravity
approximation to string theory.
In this limit 
 one finds the maximally supersymmetric 5--dimensional supergravity
with gauged $SU(4)$ symmetry~\cite{sugra}--\cite{sugrafinal}
 together with the Kaluza--Klein modes for
the ``internal'' $S_5$. There is a map~\cite{witten} between elementary
fields in the supergravity theory and gauge invariant composite operators
of the boundary $\N=4$ $SU(N)$
 SYM  theory. This theory
has an $SU(4)$ flavor symmetry which is part of its  $\N=4$ superconformal
algebra. Correlation functions of the composite operators 
in the large $N$ limit with $g_{YM}^2 N$ fixed but large 
are given by certain
classical amplitudes in supergravity.

To describe the conjecture for correlators in more detail, we note that
correlators of the $\N=4$  $SU(N)$ SYM theory are conformally related to those 
on the 4--sphere which is the boundary of (Euclidean) $AdS_5$.
Consider an operator ${\cal O}(\vec x)$ of the boundary theory, coupled to a source
$\phi_0( \vec x)$ ($\vec x$ is a point on the boundary $S_4$), and let 
$e^{-W[\phi_0]}$ denote the generating functional for correlators
of ${\cal O}(x)$. Suppose $\phi(z)$ is the field of the interior supergravity
theory which corresponds to ${\cal O}(\vec x)$ in the operator map.
Propagators $K(z,\vec x)$ between the bulk point $z$ and the
boundary point $ \vec x$ can be defined and used to construct a perturbative
solution of the classical supergravity field equation for $\phi(z)$ which is
determined by the boundary data  $\phi_0( \vec x)$. Let $S_{cl}[\phi]$ denote
the value of the
supergravity action for the field configuration $\phi(z)$.
Then the conjecture~\cite{polyakov,witten} 
is precisely that $W[\phi_0]=S_{cl}[\phi]$.
This leads to a graphical algorithm, see Fig.1, involving
 $AdS_5$ propagators and interaction vertices determined
by the classical supergravity Lagrangian. Each vertex entails a 5--dimensional
integral over $AdS_5$.

Actually, the prescriptions of \cite{polyakov} and \cite{witten}
are somewhat different. In the first~\cite{polyakov}, solutions
$\phi(z)$ of the supergravity theory satisfy a Dirichlet condition
with boundary data $\phi_0(\vec x)$ on a sphere of radius $R$ equal
to the $AdS$ length scale. In the second method~\cite{witten},
it is the infinite boundary of (Euclidean) $AdS$ space which is relevant.
Massless scalar and gauge fields satisfy Dirichlet boundary conditions
at infinity, but fields with  $AdS$ mass different from zero scale
near the boundary like $\phi(z) \rightarrow z_0^{d- \Delta} \phi_0(\vec x)$
where $z_0$ is a coordinate in the direction perpendicular to the boundary
and $\Delta$ is the dimension of the corresponding
operator $O(\vec x)$. This is
explained in detail below. Our methods apply readily only to the prescription
of \cite{witten}, although for 2--point functions we will be led  to 
consider a prescription similar to \cite{polyakov}.

To our knowledge, results for the correlators presented so far include only 
2--point functions~\cite{polyakov,witten}~\cite{sfetsos}\footnote{Very recently,
a paper has appeared~\cite{canadians} which computes special cases
of 3-- and 4--point functions of scalar operators.},
 and the purpose of the present
paper is to propose a method to calculate multi--point correlators
and present specific applications to 3--point correlators of various 
scalar composite operators and the flavor currents $J_{i}^a$ of the
boundary gauge theory. Our calculations provide explicit formulas for 
$AdS_{d+1}$ integrals needed to evaluate  generic supergravity 3--point
amplitudes involving gauge fields and scalar fields of arbitrary mass.
Integrals are evaluated for $AdS_{d+1}$, for general dimension,
to facilitate future applications of our results.
The method uses conformal symmetry to simplify
the integrand, so that the internal $(d+1)$--dimensional integral can be
simply done. This technique, which uses a simultaneous
inversion of external coordinates and external points, has been
applied to many two--loop Feynman integrals of flat four--dimensional 
theories~\cite{bakerjohnson,anselmi,erlich}. The method works well
in four flat dimensions, although there are difficulties for gauge fields, 
which arise because the invariant action $F_{\mu \nu}^2$ is inversion 
symmetric but the gauge--fixing term is not~\cite{bakerjohnson}. It is
a nice surprise that it  works even better in $AdS$ because  the
inversion is an isometry, and not merely a conformal isometry as in flat
space. Thus the method works perfectly for massive fields and for gauge
interactions in $AdS_{d+1}$ for any dimension $d$.

It is well--known that conformal symmetry severely restricts 
the tensor form of 2-- and
3--point correlation functions and frequently
determines these tensors uniquely up to a constant multiple. (For a recent
discussion, see~\cite{osborn}). This simplifies the study of the 3--point
functions. 

One of the issues we are concerned with are Ward 
identities that relate 3--point correlators with one or more
currents to 2--point
functions. It was a surprise to us this requires a minor 
 modification of the prescription of
\cite{witten} for the computation of $\< {\cal O}^I{\cal O}^J \>$
for gauge--invariant composite scalar operators.


\begin{figure} 
\begin{center} \PSbox{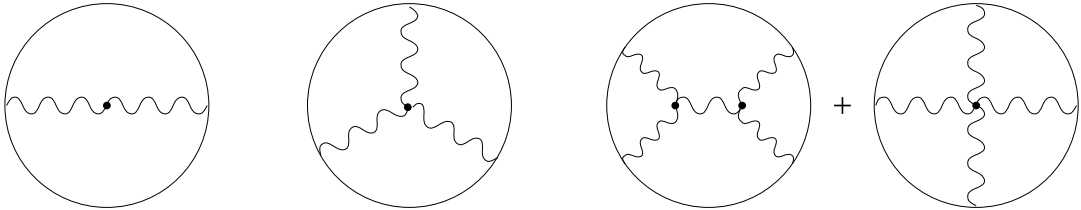 hscale=70 vscale=70}{5.6in}{1in} \end{center}
\caption{Witten diagrams.}
\end{figure}


It is also the case that some  of the correlators
we study obey superconformal non--renormalization theorems, so that the
coefficients of the conformal tensors are determined by the free--field
content of the $\N=4 $ theory and are not corrected by interactions.
The evaluation of $n$--point correlators, for $n \geq 4$, contains
more information about large $N$ dynamics, and they are given by more difficult
integrals in the supergravity construction. We hope, but cannot promise, that
our conformal techniques will be helpful here. The integrals encountered
also appear well--suited to Feynman parameter techniques, so traditional
methods may also apply. In practice, the inversion 
method reduces the number of denominators
in an amplitude, and we do apply standard Feynman parameter techniques
to the ``reduced amplitude'' which appears after inversion of coordinates.

\section{Scalar amplitudes}

It is simplest to work~\cite{witten} in the Euclidean  continuation of
$AdS_{d+1}$ which is the $Y_{-1} > 0$ sheet of the hyperboloid:
\be
\label{hyperboloid}
-(Y_{-1})^2 + (Y_{0})^2 + \sum_{i=1}^{d} (Y_i)^2 = - \frac{1}{a^2}
\ee
which has curvature $R=-d(d+1) a^2$. The change of coordinates:
\begin{eqnarray}
\label{changeofcoord}
z_i&=&\frac{Y_i}{a (Y_0 +Y_{-1})} \\
z_0&=&\frac{1}{a^2 (Y_0 +Y_{-1})} \nonumber 
\end{eqnarray}
brings the induced metric to the form of the Lobaschevsky upper half--space:
\beq
\label{lob}
ds^2   = 
\frac{1}{a^2 z_0^2} \left(  \sum_{\mu=0}^{d} dz_{\mu}^2 \right) 
=\frac{1}{a^2 z_0^2} \left( dz_0^2 + \sum_{i=1}^{d} dz_{i}^2 \right) =
 \frac{1}{a^2 z_0^2} \left( dz_0^2 +  d\vec{z}^2 \right) \nonumber
\eeq
We henceforth set $a \equiv 1$. One can verify that
the inversion: 
\beq
z'_{\mu} = \frac{z_{\mu}}{z^2}
\eeq
is an isometry
of (\ref{lob}). Its Jacobian:
\bea
\label{jacobian}
\frac{\partial z'_{\mu}}{\partial z_{\nu}} &=& (z')^2 
\left( \delta_{\mu \nu}
- 2 \frac{  z'_{\mu} z'_{\nu} }{(z')^2} \right) \\
&\equiv& (z')^2 J_{\mu \nu}(z') = (z')^2 J_{\mu \nu}(z) \nonumber
\eea
has negative determinant showing that it is a discrete isometry which is not a 
proper element of the $SO(d+1,1)$ group of
(\ref{hyperboloid}) and (\ref{lob}). Note that we define contractions such as
$(z')^2$ using the Euclidean metric 
$\delta_{\mu \nu}$, and we are usually indifferent to the question 
of raised or lowered
coordinate indices, {\it i.e.} $z^{\mu}=z_{\mu}$. When we need to contract
indices using the $AdS$ metric we do so explicitly, {\it e.g.}, 
$g^{\mu \nu} \partial_{\mu} \phi \, \partial_{\nu} \phi$,
with $g^{\mu \nu} = z_0^2\delta_{\mu \nu}$. 

The Jacobian tensor $J_{\mu \nu}$ obeys a number of identities that
will be very useful below. These include the pretty inversion property
\be
\label{3jid}
J_{\mu\nu}(x-y)=J_{\mu\rho}(x')J_{\rho\sigma}(x'-y')J_{\sigma\nu}(y')
\ee
and the orthogonality relation
\beq
\label{jorthogonal}
J_{\mu \nu}(x) J_{\nu \rho}(x)= \delta_{\mu \rho} 
\eeq

The (Euclidean) action of any massive scalar field
\beq
\label{scalaraction}
S[\phi] = \frac{1}{2} \int d^{d}z dz_0 \, \sqrt{g} 
\left[ g^{\mu \nu} \partial_{\mu} 
\phi \partial_{\nu} \phi + m^2 \phi^2
\right]
\eeq
is inversion invariant if $\phi(z)$ transforms as a scalar, {\it i.e.}
$\phi(z) \rightarrow \phi'(z) = \phi(z')$.
The wave equation is:
\beq
\label{waveequ}
\frac{1}{\sqrt{g}} \partial_{\mu} \left(
\sqrt{g} g^{\mu \nu} \partial_{\nu} \phi \right) - m^2 \phi = 0 
\eeq
\begin{equation}
\label{explicitwaveequ}
z_0^{d+1} \frac{\partial}{\partial z_0} \left[ z_0^{-d+1} \frac{\partial}{\partial
z_0} \phi(z_0, \vec{z})\right] + z_0^2 \frac{\partial}{\partial \vec{z}^2}
\phi(z_0, \vec{z})- m^2 \phi(z_0, \vec{z})  = 0
\eeq
A generic solution which vanishes as $z_0 \rightarrow \infty$ behaves like
$\phi(z_0, \vec{z}) \rightarrow z_0^{d - \Delta} \phi_{0}(\vec{z})$ as $z_0 
\rightarrow 0$, 
where $\Delta = \Delta_{+}$ is 
the largest root of the
indicial equation of (\ref{explicitwaveequ}), namely
$\Delta_{\pm} = \frac{1}{2} (d \pm \sqrt{d^2 + 4 m^2} ).$
Witten~\cite{witten} has constructed
a Green's function solution which explicitly realizes the relation
between the field $\phi(z_0, \vec{z})$ in the bulk and the boundary
configuration $\phi_0(\vec{x})$. The normalized bulk--to--boundary
Green's function\footnote{The special case $\Delta = \frac{d}{2}$
 corresponds to the lowest
$AdS$ mass allowed by unitarity, {\it i.e.} $m^2= - \frac{d^2}{4}$. 
In this case 
$\phi(z_0, \vec z) \rightarrow - z_0^{d \over 2} \,
{\rm ln} z_0 \, \phi_0 (\vec z)$ as $z_0 \rightarrow 0$ and
 the Green's function which gives this asymptotic behavior is
$ K_{d \over 2}
(z_0, \vec z ,\vec{x}) = \frac{\Gamma({d \over 2})}{2 \pi^{\frac{d}{2}} } 
\left( \frac{z_0}{z_0^2 + (\vec{z}-\vec{x})^2} \right)^{d \over 2}$.
All the formulas in the text assume the generic normalization 
(\ref{scalarprop}) valid for $\Delta > \frac{d}{2}$, obvious modifications
are needed for $\Delta = \frac{d}{2}$.}, for $\Delta > \frac{d}{2}$:
\beq
\label{scalarprop}
K_{\Delta}(z_0, \vec z ,\vec{x}) = 
\frac{\Gamma(\Delta)}{\pi^{\frac{d}{2}} \Gamma(\Delta - 
\frac{d}{2})} 
\left( \frac{z_0}{z_0^2 + (\vec{z}-\vec{x})^2} \right)^{\Delta}
\eeq
is 
a solution of (\ref{explicitwaveequ}) with the necessary singular behavior as
$z_0 \rightarrow 0$, namely: 
\beq
\label{rightsingular}
 z_0^{\Delta-d} K_{\Delta}(z_0, \vec z , \vec x) \rightarrow 
1 \cdot \delta(\vec z - \vec x)
\eeq
The solution of (\ref{explicitwaveequ}) is then related
to the boundary data by:
\beq
\label{construction}
\phi(z_0, \vec{z}) 
= 
\frac{\Gamma(\Delta)}{\pi^{\frac{d}{2}} \Gamma(\Delta -
\frac{d}{2})} \int d^{d} x\left( \frac{z_0}{z_0^2 + (\vec{z}-\vec{x})^2} 
\right)^{\Delta}
\phi_0(\vec{x}) 
\eeq

Note that the choice of $K_\Delta$ that we have taken is invariant under
translations in $\vec x$. This choice corresponds to working with a metric on
the boundary of the $AdS$ space that is flat $R^d$ with all curvature at 
infinity.
Thus our correlation functions will be  for ${\rm CFT}_{d}$ on $R^d$. 
Correlation function for other boundary metrics can be obtained by multiplying by the 
corresponding conformal factors.

It is vital to the ${\rm CFT}_{d}/AdS_{d+1}$ correspondence, and to our
method, that isometries in $AdS_{d+1}$ correspond to conformal isometries
in ${\rm CFT}_d$. In particular the inversion isometry of
$AdS_{d+1}$ is realized by the well--known conformal
inversion in ${\rm CFT}_{d}$. A scalar field (a scalar source from
the point of view of the boundary theory) $\phi_0(\vec{x})$
of scale dimension $\alpha$ transforms under
the inversion as $x_{i} \rightarrow x'_{i} / |\vec{x'}|^2$ as
$\phi_0(\vec{x}) \rightarrow \phi'_0(\vec{x})= |\vec{x'}|^{2 \alpha}
\phi_0(\vec{x'})$. The construction (\ref{construction})
can be used to show that a bulk scalar of mass $m^2$
 is related to boundary data
$\phi_0(\vec{x})$ with scale dimension $d-\Delta$. To see this one uses the
equalities:
\bea
\label{invprop}
d^{d}x & =  & \frac{d^{d}x^{'}}{|\vec{x}^{'}|^{2d}} \nonumber \\
\label{invprop2}
\left( \frac{z_0}{z_0^2 + (\vec{z}-\vec{x})^2} \right)^{\Delta} 
& = &
\left( \frac{z'_0}{(z'_0)^2 + (\vec{z'}-\vec{x'})^2} \right)^{\Delta} |
 \vec{x} ^{'}|^{2 \Delta}
\eea
and $ \phi'_0(\vec{x})
= |\vec{x}^{'}|^{2 (d-\Delta)} \phi_0(\vec{x'})$. We then find directly
that:
\beq
\frac{\Gamma(\Delta)}{\pi^{\frac{d}{2}} \Gamma(\Delta -  \frac{d}{2})} 
\int d^{d} x\left( \frac{z_0}{z_0^2 
+ (\vec{z}-\vec{x})^2} 
  \right)^{\Delta} \phi'_0(\vec{x}) 
 = \phi(z') 
\eeq
Thus conformal inversion of boundary data with scale dimension
$d-\Delta$ produces the inversion isometry in
$AdS_{d+1}$. In the ${\rm CFT}_{d}/AdS_{d+1}$ correspondence, 
$\phi_0(\vec{x})$ is viewed as the source for a scalar
operator ${\cal O}(\vec{x})$ of the ${\rm CFT}_{d}$. From
$\int d^{d}x {\cal O}(\vec{x}) 
\phi_0(\vec{x})$ one sees that ${\cal O}(\vec{x}) 
\rightarrow {\cal O}'(\vec{x}) = |\vec{x'}|^{2 \Delta}
{\cal O}(\vec{x'}) $ so that ${\cal O}(\vec{x})$
has scale dimension $\Delta$. 

Let us first review the computation of the 2--point correlator 
${\cal{O}}(\vec x)
{\cal{O}}(\vec y)$ for a ${\rm CFT}_d$ 
scalar operator of dimension $\Delta$~\cite{witten}. 
We assume that the kinetic term (\ref{scalaraction}) of
the corresponding field $\phi$ of  $AdS_{d+1}$ supergravity
is  multiplied by a constant
$\eta$ determined from the parent 10--dimensional theory.
We have, accounting for the $2$ Wick contractions:
\beq
\label{2pointscalars}
\<{\cal{O}}(\vec x) {\cal{O}}(\vec y)\> =
-2 \cdot \frac{\eta}{2} \int \frac{d^dz d z_0}{z_0^{d+1}} \left(
\partial_{\mu} K_{\Delta}(z, \vec x) z_0^2
\partial_{\mu} K_{\Delta}(z, \vec y) + m^2 K_{\Delta}(z, \vec x) 
K_{\Delta}(z, \vec y) 
\right)
\eeq 
We integrate by parts; the bulk term vanishes  by the free equation of motion
for $K$, and we get:
\bea
\label{2pointsbyparts}
\<{\cal{O}}(\vec x) {\cal{O}}(\vec y)\> &=&
+ \eta \,\lim_{\epsilon \rightarrow 0}
\int {d^dz }  \epsilon^{1-d} K_{\Delta}(\epsilon, \vec z, \vec x)
\left[ \frac{\partial}{\partial z_0} K_{\Delta}(z_0, \vec z, \vec y)
\right]_{z_0= \epsilon}
\\
&=& \eta \, \frac{ \Gamma[\Delta+1]}{ \pi^{d \over 2} \Gamma[\Delta- 
\frac{d}{2}]}
\frac{1}{|\vec x - \vec y|^{2 \Delta}} \nonumber
\eea  
where (\ref{rightsingular}) has been used. We warn readers that 
considerations of Ward identities will suggest a modification of this result
for $\Delta \neq d$. One indication that the
procedure above is delicate is that the $\partial_{\mu}K\partial_{\mu}K$
and $m^2 KK$ integrals in (\ref{2pointscalars}) are separately
divergent as $\epsilon \rightarrow 0$.

We are now ready to apply conformal methods to simplify the integrals
in $AdS_{d+1}$ which give 3--point scalar correlators in ${\rm CFT}_d$.  
We consider 3 scalar fields $\phi_I(z)$, $I=1,2,3$, in the supergravity
theory with masses $m_I$  and
interaction vertices of the form ${\cal L}_{1} = \phi_1
\phi_2 \phi_3$ and ${\cal L}_{2} = 
\phi_1 g^{\mu \nu} \partial_{\mu}\phi_2 \partial_{\nu} \phi_3$. The
corresponding 3--point amplitudes are:
\bea
A_1(\vec{x}, \vec{y}, \vec{z}) 
& = &- \int \frac{d^{d}w dw_0}{w_0^{d+1}}
K_{\Delta_1}(w,\vec{x}) K_{\Delta_2}(w,\vec{y}) K_{\Delta_3}(w,\vec{z})
\\
A_2(\vec{x}, \vec{y}, \vec{z}) & = & -\int \frac{d^{d}w dw_0}{w_0^{d+1}}
K_{\Delta_1}(w,\vec{x}) \partial_{\mu} K_{\Delta_2}(w,\vec{y}) w_{0}^2 
\partial_{\mu} 
K_{\Delta_3}(w,\vec{z}) 
\label{A2}
\eea
where $K_{\Delta_I}(w, \vec{x})$  is the Green function (\ref{scalarprop}).
 These correlators 
are conformally covariant and are of the form required by conformal
symmetry:
\beq
\label{conformalform}
A_i(\vec{x}, \vec{y}, \vec{z}) = \frac{a_i}{
|\vec{x}-\vec{y}|^{\Delta_1 + \Delta_2 - \Delta_3}
|\vec{y}-\vec{z}|^{\Delta_2 + \Delta_3 - \Delta_1}
|\vec{z}-\vec{x}|^{\Delta_3 + \Delta_1 - \Delta_2} }
\eeq
so the only issue is how to obtain the coefficients $a_1$, $a_2$.

The basic idea of our method is to use 
the inversion $w_{\mu}=\frac{w'_{\mu}}{{w'^2}}$ 
as a change of variables.
In order to use   
the simple inversion 
property~(\ref{invprop2}) of the propagator, we must also refer boundary
points to their inverses, {\it e.g.} $x_{i}=\frac{x'_{i}}{{x'^2}}$.
If this is done for a generic configuration of $\vec{x}$, $\vec{y}$, 
$\vec{z}$, there is nothing to be gained because the same integral
is obtained in the $w'$ variable. However, if we use translation symmetry
to place one boundary point at $0$, say $\vec{z}=0$, it turns
out that the denominator of the propagator attached to this point
drops out of the integral, essentially because the
inverted point is at $\infty$, and the integral simplifies.

Applied to $A_1(\vec{x},\vec{y},0)$, 
using (\ref{invprop}), 
these steps
immediately give:
\beq
A_1(\vec{x},\vec{y},0) =-
\frac{1}{|\vec{x}|^{2 \Delta_1}}
\frac{1}{|\vec{y}|^{2 \Delta_2}}
\frac{\Gamma(\Delta_3)}{\pi^{\frac{d}{2}} \Gamma(\Delta_3-  \frac{d}{2})} 
\int \frac{d^{d}w' \, dw'_0}{(w'_0)^{d+1}} \, K_{\Delta_1}(w', \vec{x'})
K_{\Delta_2}(w', \vec{y'}) \,(w'_0)^{\Delta_3}
\eeq
The remaining integral has two denominators, and it is easily done
by conventional Feynman parameter methods. We will encounter similar
integrals below so we record the general form:
\beq
\int_0^\infty dz_0 \int d^d\vec z {z_0^a\over [z_0^2+(\vec z - 
\vec x)^2]^b[z_0^2+(\vec z-\vec y)^2]^c}  \equiv 
I[a,b,c,d] |\vec x - \vec y |^{{1}+{a}+d-2b-2c  }
\nonumber
\eeq
\bea
\label{standardintegral} 
I[a,b,c,d]&=&{\pi^{d/2}\over 2}\frac{\Gamma[{a\over 2}+{1\over 2}]
\Gamma[b+c-{d\over 2}-{a\over 2}-{1\over 2}]   }{ \Gamma[b]\Gamma[c]}\\ 
&&{\Gamma[{1\over 2}+{a\over 2}+{d\over 2}-b]\Gamma[{1\over 2}
+{a\over 2}+{d\over 2}-c]\over \Gamma[1+a+d-b-c] } \nonumber
\ea
We thus find that $A_1(\vec x, \vec y, 0)$ has the spatial dependence:
\beq
\frac{1}{|\vec{x}|^{2 \Delta_1}
|\vec{y}|^{2 \Delta_2}
|\vec{x'}-\vec{y'}|^{(\Delta_1 + \Delta_2 - \Delta_3)} }
=
\frac{1}{|\vec{x}|^{ \Delta_1+ \Delta_3 - \Delta_2}
|\vec{y}|^{ \Delta_2 + \Delta_3 - \Delta_1}
|\vec{x}-\vec{y}|^{(\Delta_1 + \Delta_2 - \Delta_3)} }
\eeq
which agrees with~(\ref{conformalform}) after the translation $\vec x 
\rightarrow ( \vec x- \vec z)$, $\vec y \rightarrow (\vec y- \vec z)$. 
The coefficient $a_1$ is then:
\beq
\label{a1}
a_1  = -\frac{ 
\Gamma[\frac{1}{2} (\Delta_1 + \Delta_2 - \Delta_3)]
\Gamma[\frac{1}{2} (\Delta_2 + \Delta_3 - \Delta_1)]
\Gamma[\frac{1}{2} (\Delta_3 + \Delta_1 - \Delta_2)]
  }{        
2 \pi^d 
\Gamma[\Delta_1 - \frac{d}{2}]
\Gamma[\Delta_2 - \frac{d}{2}]
\Gamma[\Delta_3 - \frac{d}{2}]
} 
\Gamma[\frac{1}{2} (\Delta_1 + \Delta_2 + \Delta_3
-d) ]
\eeq
        
We now turn to the integral $A_2(\vec x, \vec y, \vec z)$ 
in
(\ref{A2}). It is convenient to set $\vec z = 0$. Since the structure
$ \partial_{\mu} K_2 w_{0}^2 \partial_{\mu} K_3$
is an invariant contraction and the inversion of the bulk point a 
is diffeomorphism,
we have, using (\ref{invprop2}):
\beq 
 \partial_{\mu} K_2(w,\vec{y}) 
w_{0}^2 \partial_{\mu} K_{\Delta_3}(w,0)= |\vec{y'}|^{2 \Delta_2}
\partial'_{\mu} K_{\Delta_2}(w',\vec{y'}) (w'_{0})^{2} \partial_{\mu} 
K_{\Delta_3}(w',0) 
\nonumber
\eeq
\beq 
  \sim    |\vec{y'}|^{2 \Delta_2} \frac{\partial}{\partial w'_0}
\left( \frac{w'_0}{(w'_0)^2 + (\vec{w'}-\vec{y'})^2} \right)^{\Delta_2} 
(w'_0)^2 \frac{\partial}{\partial w'_0}(w'_0)^{\Delta_3}  
\eeq
\beq 
= \Delta_2 \Delta_3 |\vec{y'}|^{2 \Delta_2} (w'_0)^{(\Delta_2 + \Delta_3)} 
\left[ \frac{1}{((w'_0)^2 + (\vec{w'}-\vec{y'})^2)^{\Delta_2}}
-\frac{2 (w'_0)^2}{((w'_0)^2 +(\vec{w'}-\vec{y'})^2  )^{\Delta_2 +1} } \right]
\nonumber
\eeq
where the normalization constants are temporarily omitted. We then find
two integrals of the form $I(a,b,c,d)$ with different parameters.
The result is:
\beq
\label{a2}
a_2= a_1 \left[ \Delta_2 \Delta_3 + \frac{1}{2} \left( 
d- \Delta_1- \Delta_2 - \Delta_3 \right)
\left(\Delta_2 + \Delta_3 - \Delta_1 \right) \right]
\eeq

As described by Witten~\cite{witten}, massive $AdS_5$ scalars are  sources
of various composite gauge--invariant scalar operators of the
$\N =4$ SYM theory. The values of the 3--point correlators of these operators
can be obtained by combining our amplitudes 
$A_1(\vec x, \vec y, \vec z)$ and  $A_2(\vec x, \vec y, \vec z)$ 
weighted by appropriate couplings from the gauged supergravity
Lagrangian.

\section{Flavor current correlators}

\subsection{Review of field theory results}

{\renewcommand{\mu}{i}
\renewcommand{\nu}{j}
\renewcommand{\rho}{k}
\renewcommand{\sigma}{l}
\renewcommand{\tau}{m}
We first review the conformal structure of the correlators $\<J_{\mu}^a(  x)
J_{\nu}^b( y)\>$ and $\<J_{\mu}^a( x)J_{\nu}^b( y)J_{\rho}^c( z)\>$ and their
non-renormalization theorems.\footnote{In this subsection,
$x$, $y$, $z$ always indicate $d$--dimensional vectors in flat
$d$--dimensional Euclidean space--time.} The situation is best understood in
4-dimensions, so we mostly limit our discussion to this physically
relevant case. The needed information probably appears in many places,
but we shall use the reference best known to us~\cite{dzfgrignani}.
 Conserved currents $J_\mu^a(x)$ have dimension $d-1$,
and transform under the inversion as
$J_\mu^a(x)\rightarrow({x'}^2)^{(d-1)}
J_{\mu \nu}({x'})J_\nu^a({x'})$.
The two-point function must take the inversion covariant, gauge--invariant form
\bea
\label{jjstructure}
\< J^a_\mu(x)J^b_\nu(y) \> \, & = & \, B \, \delta^{ab} \,{2(d-1)(d-2)\over
(2\pi)^d}{J_{\mu\nu}(x-y)\over (x-y)^{2(d-1)}} \\
& = &B\, {\delta^{ab}\over
(2\pi)^d} (\Box\delta_{\mu\nu}-\partial_\mu\partial_\nu){1\over(x-y)^{2(d-2)}}
\nonumber
\eea
where $B$ is a positive  constant, the central charge of the $J(x)J(y)$
OPE. 

In 4 dimensions the 3--point function has normal and abnormal parity
parts which we denote by $\< J^a_\mu(x)J^b_\nu(y)J^c_\rho(z) \>_\pm.$ It
is an old result~\cite{schreier} that the normal parity part is a
superposition of two possible conformal tensors (extensively studied
in~\cite{dzfgrignani}), namely
\be
\< J^a_\mu(x)J^b_\nu(y)J^c_\rho(z) \>_+=f^{abc}(k_1 D_{\mu\nu\rho}^{\rm
sym}(x,y,z) + k_2 C_{\mu\nu\rho}^{\rm sym}(x,y,z)),
\ee
where $D_{\mu\nu\rho}^{\rm sym}(x,y,z)$ and $C_{\mu\nu\rho}^{\rm
sym}(x,y,z)$ are permutation--odd tensor functions, obtained from the
specific tensors
\bea
\label{D}
D_{\mu\nu\rho}(x,y,z)& =&{1\over (x-y)^2(z-y)^2(x-z)^2}{\partial\over
\partial x_\mu}{\partial\over\partial y_\nu}\log{(x-y)^2}
{\partial\over\partial z_\rho}\log{\left({(x-z)^2\over(y-z)^2}\right)}
\\
\label{C}
C_{\mu\nu\rho}(x,y,z)&=&{1\over (x-y)^4}{\partial\over
\partial x_\mu}{\partial\over\partial z_l}\log{(x-z)^2}
{\partial\over\partial y_\nu}{\partial\over\partial z_l}\log{(y-z)^2}
{\partial\over\partial z_\rho}\log{\left({(x-z)^2\over(y-z)^2}\right)}
\nonumber
\eea
by adding cyclic permutations
\ba
D_{\mu\nu\rho}^{\rm sym}(x,y,z)=D_{i j k}(x,y,z)+D_{j k i }(y,z,x)+
D_{k i j}(z,x,y) \\
C_{\mu\nu\rho}^{\rm sym}(x,y,z)=C_{\mu\nu\rho}(x,y,z)+C_{j k i}(y,z,x)+
C_{k i j}(z,x,y).
\nonumber
\ea

Both symmetrized tensors are conserved for separated points (but the
individual permutations are not);  ${\partial\over\partial
z_\rho}D_{\mu\nu\rho}^{\rm sym}(x,y,z)$ has the local $\delta^4(x-z)$
and $\delta^4(y-z)$ terms expected from the standard Ward identity
relating 2- and 3-point correlators, while ${\partial\over\partial
z_\rho}C_{\mu\nu\rho}^{\rm sym}(x,y,z)=0$  even locally. Thus the Ward
identity implies $k_1={B\over 16\pi^6},$ while $k_2$ is an independent
constant. The symmetrized tensors are characterized by relatively
simple forms in the limit that one coordinate, say $y$, tends to
infinity:
\ba
\label{CDinfinity}
D_{\mu\nu\rho}^{\rm sym}(x,y,0)\stackrel{\textstyle\longrightarrow}
{\scriptscriptstyle y\rightarrow\infty}
{\frac{-4}{ y^6 x^4}}J_{\nu\sigma}(y)\left\{ \delta_{\mu\rho}x_\sigma-
\delta_{\mu\sigma}x_\rho-\delta_{\rho\sigma}x_\mu-2{x_\mu x_\nu
x_\sigma \over x^2}\right\}   \\
C_{\mu\nu\rho}^{\rm sym}(x,y,0)\stackrel{\textstyle\longrightarrow}
{\scriptscriptstyle y\rightarrow\infty}
\frac{8}{y^6 x^4}J_{\nu\sigma}(y)\left\{ \delta_{\mu\rho}x_\sigma-
\delta_{\mu\sigma}x_\rho-\delta_{\rho\sigma}x_\mu+4{x_\mu x_\nu
x_\sigma \over x^2}\right\} \nonumber
\ea

In a superconformal--invariant theory with a fixed line parametrized by
the gauge coupling, such as ${\cal N}=4$ SYM theory, the constant $B$
is exactly determined by the free field content of the theory, {\it i.e.} 1--loop
graphs. This is the non-renormalization theorem for flavor central
charges proved in~\cite{nonrenorm}. The argument is quite simple. The fixed
point value of the central charge is equal to the external trace
anomaly of the theory with source for the currents~\cite{anselmialone,anselmi}. 
Global ${\cal
N}=1$ supersymmetry relates the trace anomaly to the R-current anomaly,
specifically to the $U(1)_RF^2$ ($F$ is for flavor) which is one-loop
exact in a conformal theory. Its value depends on the r--charges
and the flavour quantum numbers of the fermions of the theory, and it
is independent of the couplings. For an $\N=1$ theory with chiral superfields
$\Phi^i$ with (anomaly--free) r--charges
 $r_i$ in irreducible representations $R_i$ of
the gauge group, the fixed point value of the central charge was given
in (2.28) of~\cite{positivity} as
\be
B\delta^{ab}=3\sum_i({\rm dim} R_i)(1-r_i){\rm Tr}_i(T^aT^b).
\ee
For ${\cal N}=4$ SYM we can restrict to the $SU(3)$ subgroup of the
full $SU(4)$ flavour group that is manifest in an ${\cal N}=1$
description. There is a triplet of $SU(N)$ adjoint $\Phi^i$ with
$r={2\over 3}.$ We thus obtain
\be\label{B}
B=3(N^2-1){1\over 3}\cdot{1\over 2}={1\over 2}(N^2-1).
\ee

We might now look forward to the $ AdS_5$ calculation with the
expectation that the value  found for $k_1$ will be determined by the
non--renormalization theorem, but $k_2$ will depend on the large $N$
dynamics and differ from the free field value.
  Actual results will force us to revise this
intuition. We now discuss the 1--loop contributions in the field theory
and obtain the values of $k_1$ and $k_2$ for later comparison with
$ AdS_5$. 

Spinor and scalar 1-loop graphs were expressed as linear combinations
of $D^{\rm sym}$ and $C^{\rm sym}$ in~\cite{dzfgrignani}. For a single
$SU(3)$ triplet of left handed fermions and a single triplet of
complex bosons one finds
\ba
\< J^a_\mu(x)J^b_\nu(y)J^c_\rho(z) \>_+^{\rm fermi}={4\over 3}
{f^{abc}\over (4\pi^2)^3}( D_{\mu\nu\rho}^{\rm
sym}(x,y,z) -{1\over 4} C_{\mu\nu\rho}^{\rm sym}(x,y,z)) \\
\<J^a_\mu(x)J^b_\nu(y)J^c_\rho(z)\>^{\rm bose}={2\over 3}
{f^{abc}\over (4\pi^2)^3}( D_{\mu\nu\rho}^{\rm
sym}(x,y,z) +{1\over 8} C_{\mu\nu\rho}^{\rm sym}(x,y,z))
\nonumber
\ea
The sum of these, multiplied by $N^2-1$ is the  total 1-loop result in the
${\cal N}=4$ theory:
\be
\label{jjjfieldresult}
\<J^a_\mu(x)J^b_\nu(y)J^c_\rho(z)\>_+^{\rm {\cal N}=4}=
{(N^2-1)f^{abc}\over 32\pi^6}( D_{\mu\nu\rho}^{\rm
sym}(x,y,z) -{1\over 8} C_{\mu\nu\rho}^{\rm sym}(x,y,z)).
\ee
We observe the agreement with the value of $B$ in (\ref{B}) and the
fact that the 
free field ratio of $C^{\rm sym}$ and $D^{\rm sym}$ tensors is
$-{1\over 8}.$

Since the $SU(4)$ flavor symmetry is chiral, the 3--point current correlator
also has an abnormal parity part 
$\<J^a_{\mu}J^b_{\nu}J^c_{\rho}\>_{-}$.
It is well--known that there is a unique conformal 
tensor--amplitude~\cite{schreier} in this section, which is a constant
multiple of the fermion triangle amplitude, namely
\beq
\label{anomalyfeynman}
\< J^a_{\mu}(x)J^b_{\nu}(y)J^c_{\rho}(z)\>_{-}
= - \frac{N^2 -1}{32 \pi^6}i d^{a b c}
\frac{{\rm Tr} \left[ \gamma_5 \gamma_{\mu} (\not{x} - \not{y})
\gamma_{\nu} (\not{y} - \not{z}) \gamma_{\rho} (\not{z} - \not{x})\right]}{ 
(x-y)^4 (y-z)^4 (z-x)^4}
\eeq
where the $SU(N)$ $f$ and $d$ symbols are defined by ${\rm Tr}( T^a
T^b T^c) \equiv \frac{1}{4}(i f^{a b c } + d^{a b c })$ with $T^a$ hermitian
generators normalized as ${\rm Tr} T^a T^b = \frac{1}{2} \delta^{ab}$. 
The coefficient is again
``protected''  by a non--renormalization theorem, namely the Adler--Bardeen
theorem (which is independent of SUSY and conformal symmetry). After
bose--symmetric regularization~\cite{erlich} of the short
distance singularity, one finds the anomaly
\beq
\label{anomalypositionspace}
\frac{\partial}{\partial z_{k}} 
\<J^a_{i }(x)J^b_{j}(y)J^c_{k}(z) \>_{-}
= - \frac{N^2 -1}{48 \pi^2}i d^{a b c} 
\epsilon^{i j l m}
\frac{\partial}{\partial x_{l}}\frac{\partial}{\partial y_{m}}
\delta(x-z) \delta(y-z) 
\eeq
If we minimally couple the currents $J_i^a(x)$ to background
sources $A_i^a(x)$ by adding to the action a term
$\int d^4 x J_i^a(x) A_i^a(x)$, 
 this information can be presented as the operator equation:
\beq
\label{anomalyoperator}
(D_i J_i(z))^a =
\frac{\partial}{\partial z_{i}} J_{\mu}^a(z)
+ f^{a b c} A_i^b(z) J_i^c (z)=
 \frac{N^2 -1}{96 \pi^2} i d^{a b c} \epsilon_{ j k lm}
\partial_{\nu} (A^b_{\rho} \partial_{\sigma} A^c_{\tau} + 
\frac{1}{4} f^{c d e} A_k^b A_l^d A_m^e
)   
\eeq
where the cubic term in $A_i^a$ is determined by the Wess--Zumino
consistency conditions (see e.g. \cite{jackiw}).

}

The ${\rm CFT}_4/AdS_5$ correspondence can also be used to calculate the
large $N$ limit of correlators
$\<J_i^a( x) {\cal O}^I( y) {\cal O}^J( z)\>$ 
and $\< J_i^a( x) J_j^b( y) {\cal O}^I( z)\>$ 
where ${\cal O}^I$ is a gauge--invariant composite scalar operator
of the $\N=4$ SYM theory. For example, one can take
${\cal O}^I$ to be a $k$--th rank traceless symmetric tensor 
${\rm Tr}\, X^{\alpha_1} \cdot \cdot \cdot X^{\alpha_k}$
(the explicit subtraction of traces is not  indicated) formed from the real
scalars $X^{\alpha}$, $\alpha=1,...,6$, in the 6--dimensional
representation of $SU(4) \cong SO(6)$, and there are other
possibilities in the operator map
discussed by Witten~\cite{witten}. We will
compute the corresponding supergravity amplitudes in the next section,
and we record here the tensor form required
by conformal symmetry.

For $\< J_i^a {\cal O}^I{\cal O}^J\>$ there is a unique conformal
tensor for every dimension $d$ given by
\beq
\label{JOOtensor}
\<J_i^a( z) {\cal O}^I( x) {\cal O}^J( y)\> = 
\xi \stackrel{a \, \ }{S_{i}^{IJ}}(z, x, y)
\eeq
\beq
 \equiv  -\xi (d-2) \stackrel{a \ \,}{T^{IJ}}
\frac{1}{( x -  y)^{2 \Delta - d +2}}
\frac{1}{( x - z)^{d-2} ( y -  z)^{d-2}}
\left[
\frac{(x - z)_i}{( x -  z)^2}-
\frac{(y - z)_i}{( y -  z)^2}
\right] 
\eeq
where $\xi$ is a constant and $\stackrel{a \ \,}{T^{IJ}}$
are the Lie algebra generators. This correlator satisfies a Ward identity
which relates it to the 2--point function
$\< {\cal O}^I(x){\cal O}^J(y) \>$. Specifically:
\bea
\label{ward}
\xi \frac{\partial}{\partial z_i} \stackrel{a \, \ }{S_{i}^{IJ}}(z, x, y)
& = & \xi \frac{(d-2)2 \pi^{d \over 2}}{\Gamma[{d \over 2}]}
 \stackrel{a \ \,}{T^{IJ}}
\left(  
\delta^d(x-z) - \delta^d(y-z)
\right) \frac{1}{(x-y)^{2 \Delta} }\\
& = & \delta^d(x-z)\stackrel{a \ \,}{T^{IK}}
\< {\cal O}^K(x) {\cal O}^J(y) \> +
\delta^d(y-z)\stackrel{a \ \,}{T^{JK}}\< {\cal O}^I(x) {\cal O}^K(y) \> 
\nonumber
\eea

There is also a unique tensor form for
$\<J_i J_j  {\cal O}\>$ (we suppress group theory labels) 
which is given in~\cite{anselmi}:
\beq
\label{JJOtensor}
\<J_i( x) J_j( y) {\cal O}( z)\>
= \zeta  R_{ij}(x,y,z) \equiv
\zeta  
\frac{(6-\Delta) J_{ij}(x - y) -\Delta J_{ik}(x-z)J_{kj}(z-y)
}{(x-y)^{6-\Delta} (x-z)^{\Delta} (y - z)^{\Delta}
}
\eeq
where $\zeta$ is a constant.

\subsection{Calculations in $AdS$ supergravity}
The boundary values $A_i^a(\vec x)$ 
of the gauge potentials $A_\mu^a(x)$ of gauged supergravity are the
sources for the conserved flavor currents $J_i^a(\vec x)$ of the boundary
${\rm SCFT}_4$.  
It is  sufficient for our
purposes to ignore non-renormalizable $\phi^n F_{\mu\nu}^2$
interactions and represent the gauge sector of the supergravity by the
 Yang--Mills and Chern--Simons terms (the latter for $d+1=5$)
\be
\label{sugraaction}
S_{cl}[A]=\int d^dz dz_0 \left[ \sqrt{g}\, {F_{\mu\nu}^a F^{\mu\nu a}\over 4
g^2_{\small SG}}+
\frac{i k}{96 \pi^2}
 \left( d^{abc}\epsilon^{\mu\nu\lambda\rho\sigma}
A_\mu^a\partial_\nu A^b_\lambda\partial_\rho A^c_\sigma  
+ \cdot \cdot \cdot \right)\right]
\ee
The coefficient $\frac{k}{96 \pi^2}$,
where $k$ is an integer, is
the correct normalization factor 
for the 5--dimensional Chern--Simons term ensuring that under a 
large gauge transformation
the action changes by an unobservable phase
$2 \pi i n$ (see e.g. \cite{jackiw}).
The couplings $g_{SG}$ and $k$ could in principle be
determined from dimensional reduction of  the parent 10 dimensional
theory, but we shall ignore this here. Instead, they
will be fixed in terms of  current correlators of the
boundary theory which are exactly known because they satisfy
 non-renormalization theorems.

To obtain flavor--current correlators in the boundary CFT from $AdS$ 
supergravity, we need a Green's function $G_{\mu i}(z, \vec x)$
to construct the gauge potential $A_{\mu}^a(z)$ in the bulk from its
boundary values $A_i^a(\vec x)$. We will work in $d$ dimensions.
There is the gauge freedom to redefine $G_{\mu i}(z, \vec x) 
\rightarrow G_{\mu i}(z, \vec x) +\frac{\partial}{\partial z_{\mu}} 
\Lambda_i(z, \vec x)$
which leaves boundary amplitudes obtained from the action~(\ref{sugraaction})
invariant. Our method requires a conformal--covariant
propagator, namely
\bea
\label{G}
G_{\mu i}(z, \vec x) 
& =  & C^{d} \frac{ 
z_0^{d-2}}{ [z_0^2 +(\vec z - \vec x)^2]^{d-1} } J_{\mu i}(z - \vec x)\\
& = & C^{d} \left( \frac{z_0}{( z - \vec x)^2} 
\right)^{d-2} \partial_{\mu} \left( \frac{(z - \vec x)_i}{(z - \vec x)^2 }
\right) \label{Gform}
\eea
which satisfies the gauge field equations of motion in the bulk variable $z$.
The normalization constant $C^d$ is determined by requiring
 that as $z_0 \rightarrow 0$, 
$G_{ji}(z, \vec x) \rightarrow 1 \cdot \delta_{j i} \delta({\vec x})$: 
\beq
\label{Cd}
C^{d} = \frac{ \Gamma(d)}{ 2 \,\pi^{d \over 2} \Gamma({d \over 2})}
\eeq
This Green's function does not satisfy boundary transversality
({\it i.e.} $\frac{\partial}{\partial x_i} G_{\mu i}(z, \vec x) = 0$),
but the following gauge--related propagator does\footnote{For even $d$,
the hypergeometric function in (\ref{gbar}) is actually a rational
function. For instance for 
$d=4$, $\bar{G}_{\mu i}(z, \vec x) =G_{\mu i}(z, \vec x) + 
  \frac{\partial}{\partial z_{\mu}}\left\{ \frac{C^d}{12} 
\frac{\partial}{\partial z_i}
  \left( \frac{2 z_0^2 +(\vec z - \vec x)^2}{[z_0^2 + (\vec z - \vec x)^2]^2}
\right)
\right\}.$}: 
\beq
\label{gbar}
\bar{G}_{\mu i}(z, \vec x) =G_{\mu i}(z, \vec x) + 
  \frac{\partial}{\partial z_{\mu}}\left\{ \frac{C^d z_0^{2-d}}{(d-2)(d-1)
(\Gamma[\frac{d}{2}])^2 } 
\frac{\partial}{\partial z_i}
F \left[ d-1, \frac{d}{2} -1, \frac{d}{2}; -\frac{(\vec z - \vec x)^2}{z_0^2}
\right]
\right\} 
\eeq
(Both $G_{\mu i}(z, \vec x)$ and $\bar{G}_{\mu i}(z, \vec x)$
differ by gauge terms from the Green's function used by Witten~\cite{witten}).
The gauge equivalence of inversion--covariant and transverse propagators
ensures that the method produces boundary current 
 correlators which are conserved.

Notice that in terms of the conformal tensors $J_{\mu i}$ the abelian
field strength 
 made from the Green's function
takes a remarkably simple form:
\beq
\label{dG}
\partial_{[\mu} G_{\nu] i}(z, \vec x) 
= 
(d-2) C^{d}  \frac{ 
z_0^{d-3}}{ [z_0^2 +(\vec z - \vec x)^2]^{d-1} } 
J_{0 [\mu}(z - \vec x) J_{\nu] i}(z - \vec x) 
\eeq 
as  easily checked by using for
$G_{\mu i}$ the representation (\ref{Gform}). 

We stress again that the inversion $z_{\mu}=z'_{\mu}/(z')^2$ 
is a coordinate
transformation which is an isometry of $AdS_{d+1}$. It acts as a 
diffeomorphism on the internal indices $\mu, \nu, \dots$
of $G_{\mu i}, G_{\nu j}, \dots$.
Since these indices are covariantly contracted at an internal point
$z$, much of the algebra required to change integration variables can
be avoided. The inversion $\vec x = \vec{x'}/ (\vec{x'})^2$ 
of boundary points
is a conformal isometry which 
acts on the external index $i$
and also changes the Green's 
function by a conformal factor. Thus the change of variables amounts
to the replacement:
\bea
\label{Ginversion}
G_{\mu i}(z ,\vec x) &= & z'^2 J_{\mu \nu}(z') \cdot 
(\vec{x'})^2  J_{ki}(\vec x') \cdot (\vec{x'})^{2(d-2)}
\, C^d \, \frac{ (z'_0)^{d-2} J_{\nu k} (z' -\vec{x'}) }{[(z'_0)^2  
+ (\vec{z'}- \vec{x'})^2]^{d-1}} \\
& = & {\partial z'_{\nu} \over \partial z_{\mu}}
\, \cdot 
{\partial x'_{k} \over \partial x_{i}}\, \cdot 
(\vec{x'})^{2(d-2)}
G_{\nu k} (z', \vec{x'}) \nonumber \\
& = & {\partial z'_{\nu} \over \partial z_{\mu}}
\, \cdot 
{\partial x'_{k} \over \partial x_{i}}\, \cdot G'_{\nu k} (z', \vec{x'})
\nonumber
\eea
$\partial_{[\mu} G_{\nu] i}(z, \vec x)$ 
will also transform  conformal--covariantly under inversion 
(compare equ.(\ref{Ginversion})):
\beq
\label{dGinv}
\partial_{[ \mu} G_{\nu] i}(\vec x, z) 
=
(z')^2 J_{\mu \rho}(z') \cdot (z')^2 J_{\nu \sigma}(z') \cdot
(\vec{x'})^2  J_{ki}(\vec x') \cdot (\vec{x'})^{2(d-2)}
\partial'_{[ \rho} G_{\sigma] k}(\vec{x'}, z')  
\eeq
as one can directly check from (\ref{dG}) using the identity
(\ref{3jid}).

${\bf \<J_i^a J_j^b\>}$:
To obtain the current--current correlator we follow the same 
procedure~\cite{witten}
as for the scalar 2--point function, 
eq.(\ref{2pointscalars}--\ref{2pointsbyparts}):
\bea
\<J_i^a(\vec x) J_j^b(\vec y)\> & = &
-\delta^{a b}\, 2 \cdot \frac{1}{4 g_{SG}^2} \int \frac{d^dz d z_0}{z_0^{d+1}}
\, \partial_{[\mu} G_{\nu]i} (z, \vec x)\, z_0^4 \,
\partial_{[\mu} G_{\nu]j}(z,\vec y) \nonumber \\
 &= & 
+\frac{\delta^{a b}  }{2 g_{SG}^2} \lim_{\epsilon \rightarrow 0}
\int d^dz \, \epsilon^{3-d}\, 
\, 2 G_{\nu i} (\epsilon, \vec z, \vec x) \, 
\left[ \partial_{[0} G_{\nu ] j}(z_0, \vec z, \vec y)
 \right]_{z_0= \epsilon} \nonumber \\
\label{jjsugra}
&= &\delta^{a b} \,
\frac{C^d (d-2)}{g_{SG}^2} \, 
\frac{J_{i j}(\vec x - \vec y)}{|\vec x - \vec y|^{2(d-1)}}
\eea
 which is of the form~(\ref{jjstructure}) with  
$B=\frac{1}{g_{SG}^2} \frac{2^{d-2} \pi^{\frac{d}{2}} \Gamma{[d]}}{(d-1)
\Gamma[\frac{d}{2}]}$.
According to the conjecture~\cite{maldacena,polyakov,witten}, 
(\ref{jjsugra}) represents the large--$N$ value of the 2--point function
for  $g_{YM}^2 N$ fixed but large.
Let us now consider the case $d=4$. By
 the non--renormalization theorem proven in~\cite{nonrenorm},
the coefficient in (\ref{jjstructure}) is protected against
quantum corrections. Hence, at leading order in $N$, the strong--coupling
result (\ref{jjsugra}) has to match the 1--loop computation (\ref{B}).
We thus learn:
\beq
\label{gSG}
g^{d+1=5}_{SG} = \frac{4 \pi}{N}
\eeq

${\bf \< J_i^a J_j^b J_k^c\>_{+}}$:
The  vertex relevant to the computation of the normal
parity part of $<J_i^a(\vec x) J_j^b(\vec y) J_k^c(\vec z)>$ 
comes from the Yang--Mills term of the
action (\ref{sugraaction}), namely
\beq
\label{ym3vertex}
\frac{1}{2 g_{SG}^2} \int \frac{d^dw dw_0}{w_0^{d+1}} \,
i f^{a b c} \, \partial_{[\mu} A_{\nu]}^a (w)
\, w_0^4 \, A_{\mu}^b(w) A_{\nu}^c(w)
\eeq
We then have
\bea
\label{jjjFFF}
\<J_i^a(\vec x) J_j^b(\vec y) J_k^c(\vec z)\>_{+}
& = &
-\frac{i f^{a b c }}{2 g_{SG}^2} \ 2  \cdot F^{\rm sym}_{i j k}
(\vec x, \vec y, \vec z)
  \\
& \equiv  &
-\frac{i f^{a b c }}{2 g_{SG}^2} 
\   2  \left[   
 F_{i j k}(\vec x, \vec y ,\vec z)
+ F_{ j k i}( \vec y, \vec z, \vec x) 
+ F_{k i j}(\vec z, \vec x, \vec y) \right]  \nonumber
\eea
where
\beq
\label{F}
F_{i j k}(\vec x, \vec y ,\vec z) =\int \frac{d^dw dw_0}{w_0^{d+1}} \,
\partial_{[\mu} G_{\nu] i}(w, \vec x) \, w_0^4 \, G_{\mu j}(w,\vec y)
 G_{\nu k}(w, \vec z)
\eeq
(The extra factor of 2 in (\ref{jjjFFF}) correctly accounts for the $3 !$ Wick
contractions). To apply the method of  inversion,
it is convenient to set $\vec x =0$.
Then, changing integration variable $w_{\mu} = \frac{w'_{\mu}}{(w')^2}$ and
inverting the external points, $y_{i} = \frac{y'_{i}}{|\vec{y'}|^2}$,
$z_{i} = \frac{z'_{i}}{|\vec{z'}|^2}$, we achieve the  simplification
(using (\ref{Ginversion}),(\ref{dGinv}),(\ref{jorthogonal})):
\bea
&& F_{i j k}(0, \vec y, \vec z) = \nonumber
\\
&& =  
|\vec{y'}|^{2(d-1)} 
|\vec{z'}|^{2(d-1)}  J_{jl}(\vec{y'}) J_{km}(\vec{z'}) \nonumber \\
\label{FF'}
&& \cdot \int \frac{d^dw' dw'_0}{(w'_0)^{d+1}} \,
  \partial'_{[\mu} G_{\nu] i}(w',0) \, (w'_0)^4 \, G_{\mu l}(w,\vec{y'})
 G_{\nu m}(w', \vec{z'}) \\ 
&& =(C^d)^3  
\frac{J_{jl}(\vec{y})}{|\vec{y}|^{2(d-1)} } \,
\frac{J_{km}(\vec{z})}{|\vec{z}|^{2(d-1)} }  
  \int \frac{d^dw' dw'_0}{(w'_0)^{d+1}} \,
 \Big[ 
\partial'_{[\mu} (w'_0)^{d-2} \partial'_{\nu]}(w'_i)
 \, (w'_0)^4 \nonumber \\
&& \label{F'processed}
\qquad
\qquad
\quad
  \frac{(w'_0)^{d-2}}{(w'-\vec{y'})^{2(d-2)}}
 \frac{(w'_0)^{d-2}}{(w'-\vec{z'})^{2(d-2)}}
 J_{\mu l}(w,\vec{ y'})
 J_{\nu m}(w', \vec{ z'})  \Big]  \\
&&=(C^d)^3 
\frac{J_{jl}(\vec{y})}{|\vec{y}|^{2(d-1)} } \,
\frac{J_{km}(\vec{z})}{|\vec{z}|^{2(d-1)} }  
  \int {d^dw' dw'_0} \, 
\frac{ (d-2) (w'_0)^{2d-4}J_{l [0}(w' - \vec t) J_{i] m}( w')
}{[(w'_0)^2 +(\vec{w'}- \vec{t})^2]^{d-1}
[(w'_0)^2 +(\vec{w'})^2]^{d-1} } \nonumber 
\eea
where in the last step we have defined $\vec{t} \equiv \vec{y'} - \vec{z'}$.
Observe that in going from (\ref{F}) to (\ref{FF'}) we just had to replace the
original variables with primed ones and  pick conformal Jacobians for the 
external
(Latin) indices: the internal Jacobians nicely collapsed with each other
(recall the contraction rule (\ref{jorthogonal}) for  $J_{\mu i}$ tensors)
and with the factors  of $w'$ coming from the inverse metric.
The integrals in (\ref{F'processed}) now have  two denominators and 
through straightforward manipulations can be rewritten as derivatives with 
respect to the external coordinate $\vec t$ of standard integrals of the form
(\ref{standardintegral}). We thus obtain:
\bea
F_{i j k}(\vec x, \vec y, \vec z)& = &
- \,\frac{J_{jl}(\vec{y}-\vec x)}{|\vec{y}- \vec x|^{2(d-1)} } \,
\frac{J_{km}(\vec{z}-\vec x)}{|\vec{z}-\vec x |^{2(d-1)} }  
 (C^d)^3 \pi^{\frac{d+2}{2}} 2^{3-2d} \left( \frac{d-2}{d-1}
\right) \frac{\Gamma\left[{d \over 2}\right]}{\left[ \Gamma\left[
\frac{d+1}{2}\right] \right]^2}
 \nonumber \\
\label{F't}
&&  \cdot \frac{1}{|\vec t|^d}
\left[ 
\delta_{lm} t_i + (d-1)\delta_{il} t_m +(d-1) \delta_{im} t_l
-d \frac{t_i t_l t_m}{|\vec{t}|^2}
\right] 
\eea 
where we have restored the $\vec x$ dependence, so that now
$\vec t \equiv (\vec y - \vec x)' - (\vec z - \vec x)'$.
We now  add permutations to 
obtain 
$F^{\rm sym}_{i j k}
(\vec x, \vec y, \vec z)$ in (\ref{jjjFFF}). The 
final step is to express $F^{\rm sym}_{i j k}$ as a linear combination
of the conformal tensors $D^{{\rm sym}}_{i j k}$
 and $C^{{\rm sym}}_{i j k}$ of Section 3.1. It is simplest,
and by conformal invariance not less general, to work
in the special configuration $\vec z =0$ and
$|\vec y| \rightarrow \infty$. After careful algebra
we obtain
\bea
F^{\rm sym}_{i j k}(\vec x, |\vec y| \rightarrow \infty, 0)
&= &- \,  
 (C^d)^3 \pi^{\frac{d+2}{2}} 2^{2-2d} (2d-3)\left( \frac{d-2}{d-1}
\right) \frac{\Gamma\left[{d \over 2}\right]}{\left[ \Gamma\left[
\frac{d+1}{2}\right] \right]^2} 
\\
&& \cdot \frac{J_{jl}(\vec y)}{|\vec{y}|^{2(d-1)} 
|\vec x |^{ d}}
\left\{ \delta_{ik} x_l-
\delta_{il}x_k-\delta_{kl}x_i- \frac{d}{2d-3} {x_i x_j
x_l \over x^2}\right\} \nonumber
\eea
Now take $d =4$; comparison with (\ref{CDinfinity}) gives
\beq
\label{FsymCD}
F^{\rm sym}_{i j k}(\vec x, \vec y , \vec z) =
\frac{1}{ \pi^4 } 
\left(
D^{{\rm sym}}_{i j k}(\vec x, \vec y, \vec z)
-\frac{1}{8}C^{{\rm sym}}_{i j k}(\vec x, \vec y, \vec z)
\right)
\eeq
and finally, from (\ref{jjjFFF}) and (\ref{gSG}):
\bea
\label{jjjfinal}
\<J_i^a(\vec x) J_j^b(\vec y) J_k^c(\vec z)\>_{+}& =&
\frac{ f^{a b c}}{2 \pi^4 g_{SG}^2}
\left(
D^{{\rm sym}}_{i j k}(\vec x, \vec y, \vec z)
-\frac{1}{8}C^{{\rm sym}}_{i j k}(\vec x, \vec y, \vec z)
\right) \\
\nonumber
&=& \frac{N^2 \,  f^{abc}}{32 \pi^6}
\left(
D^{{\rm sym}}_{i j k}(\vec x, \vec y, \vec z)
-\frac{1}{8}C^{{\rm sym}}_{i j k}(\vec x, \vec y, \vec z)
\right)
\eea
which, at  leading order in $N$, precisely agrees with
the 1--loop result (\ref{jjjfieldresult}). 

The correlator (\ref{jjjfinal})
calculated from $AdS_5$ supergravity is supposed
to reflect the strong--coupling dynamics of the $\N =4$ SYM
theory at  large $N$. The exact agreement found with the free--field
result therefore requires some comment. As discussed in Section 3.1, the 
coefficient of the $D$ tensor is fixed by the Ward identity that
relates it to the constant $B$ in the 2--point function,
and we matched the latter to the 1--loop result by a non--renormalization
theorem. So agreement here is just a check that we have done the integral
correctly. However, the fact that the ratio of the $C$ and $D$ tensors
coefficients also agrees with the free field value was initially
a surprise. Upon further thought, we see that our argument that the value of
$k_2$ was a free parameter used only $\N=0$ conformal symmetry,
and superconformal symmetry may impose some constraint. Indeed,
in an $\N=1$ description of the $\N=4$ SYM theory, we have the flavor
$SU(3)$ triplet $\Phi^i$ of ($SU(N)$ adjoint) chiral superfields, 
together with their
adjoints $\bar{\Phi}^i$. 
The $SU(3)$ flavor currents are the $\bar{\theta} \theta$
components of composite scalar superfields $K^a(\vec x, \theta, \bar{\theta})
= {\rm Tr} \, \bar{\Phi} T^a \Phi$, where $T^a$ is a fundamental $SU(3)$
matrix. Just as $\N = 0$ conformal invariance
constrains the tensor form of 2-- and 3--point correlators, $\N =1$
superconformal symmetry will constrain the superfield
correlators $\<K^a K^b\>$ and
$\<K^a K^b K^c\>$. We are not aware of a specific analysis, but
it seems likely~\cite{osbornprivate} that there are
only two possible superconformal
amplitudes for $\<K^a K^b K^c\>$, one proportional to $f^{abc}$
and the other to $d^{a bc}$. The $f^{abc}$ amplitude contains
the normal parity $\<J_i^a J_j^b J_k^c\>_{+}$ in its 
$\theta$--expansion,
and this would imply that the ratio $-\frac{1}{8}$ of the coefficients
of the $C$ and $D$ tensors must hold in any $\N=1$ superconformal theory.

${\bf \<J_i^a J_j^b J_k^c\>_{-}}$ Witten~\cite{witten} has sketched
an elegant argument that allows to  read the value
of the abnormal parity part
of 
the 3--current correlator 
directly from the supergravity action (\ref{sugraaction}),
with no integral to compute.
Under an infinitesimal gauge transformation of the bulk gauge potentials,
$\delta_{\Lambda} A_{\mu}^a = (D_{\mu} \Lambda)^a$, the
variation of the the action is purely a boundary term coming
from the Chern--Simons 5--form:
\beq
\label{chernvariation}
\delta_{\Lambda} \,S_{cl}
= \int d^4z  \, \Lambda ^a(\vec z) \,
\left(-\frac{ik}{96 \pi^2}\right) 
d^{a b c} \epsilon^{i j k l}
\partial_{i} (A^b_{j} \partial_{k} A^c_{l} + 
\frac{1}{4} f^{c d e} A_j^b A_k^d A_l^e
)   
\eeq
By the conjecture~\cite{maldacena,polyakov,witten}, $S_{cl}[A^a_{\mu}(z)]=
W[A_i^a(\vec z)]$, the generating functional for current
correlators in the boundary theory. Since by construction
$J^a_i(\vec x) =  \frac{\delta W[A]}{\delta A_i^a(\vec x)}$, one has:
\beq
\label{deltaSdeltaW}
\delta_{\Lambda} \,S_{cl}[A_{\mu}^a(z)] =
\delta_{\Lambda} \,W[A_{\mu}^a(\vec z)] 
=\int d^4 z [D_i \Lambda(\vec z)]^a J^a_i(\vec z) =
- \int d^4 z \Lambda^a(\vec z) [D_i J_i(\vec x)]^a
\eeq
and comparison with (\ref{chernvariation}) gives
\beq
\label{anomalyfromads}
(D_i J_i(\vec z))^a =
 \frac{i k}{96 \pi^2}  d^{a b c} \epsilon^{ j k l m}
\partial_{j} (A^b_{k} \partial_{l} A^c_{m} + 
\frac{1}{4} f^{c d e} A_j^b A_k^d A_l^e)
\eeq
which has precisely the structure (\ref{anomalyoperator}). Thus
the ${\rm CFT}_4/AdS_5$ correspondence gives a very
concrete physical realization of the well--known
mathematical relation between the gauge anomaly in $d$ dimensions
and the gauge variation of a $(d+1)$--dimensional Chern--Simons
form. Witten~\cite{witten} has argued that (\ref{anomalyfromads})
is an exact statement even at finite $N$ (string--loop effects)
and for finite 't Hooft coupling $g^2_{YM} N$
(string corrections to the classical supergravity action), 
which  is of course what one expects from the Adler--Bardeen theorem.
Matching (\ref{anomalyfromads}) with the 1--loop result
(\ref{anomalyoperator}) we are thus led to
identify $k= N^2 -1$.

${\bf \<J_i^a J_j^b { \bf \cal O}\>}$:
 The next 3--point correlator
to be discussed is $\<J_i^a(\vec x) J_j^b(\vec y) {  \cal O}^I
(\vec z) \>$. For this purpose we suppress  group
 indices and consider a supergravity interaction of the form
\beq
\frac{1}{4} \,
\int d^dw dw_0 \, \sqrt{g} \, g^{\mu \rho} g^{\nu \sigma}
\, \phi \, \partial_{[ \mu} A_{\nu ]}  \partial_{[ \rho} A_{\sigma ]}  
\eeq
This leads to the boundary amplitude
\beq
\frac{1}{2} \int \frac{d_d w dw_0}{w_0^{d+1}}\,
K_{\Delta}(w,\vec z) \partial_{[ \mu} G_{\nu ]}(w , \vec x) 
w_0^2 \partial_{[ \mu} G_{\nu ]}(w,\vec y)  
\eeq
We set $\vec y = 0$, apply the method of inversion and obtain the integral
\bea
T_{ij}(\vec x, 0, \vec z) 
& = &
(C^d)^2 
\frac{\Gamma[\Delta]}{\pi^{d \over 2} \Gamma[\Delta- {d \over 2}]}
\frac{(d-2) J_{i k}(\vec x)}{|\vec z|^{2 \Delta} |\vec x|^{2(d-1)}}
\\
& & \cdot
\int d^d w' dw'_0 \,
\left( \frac{w'_0}{(w'-\vec{z'})^2}
\right)^\Delta
\frac{\partial}{\partial w'_{[0}}
\left( \frac{w'_0}{(w'-\vec{z'})^2}
\right)^{d-2}
\frac{\partial}{\partial w'_{j]}}
\frac{(w' - x')_{k}}{(w' - \vec{x'})^2}
\nonumber
\eea
This can be evaluated as a fairly standard Feynman integral 
with two denominators. The result is
\beq
T_{ij}(\vec x, \vec y, \vec z)= - \frac{\Delta}{8 \pi^2}
\frac{\Gamma[\Delta]}{\pi^{d \over 2} \Gamma[\Delta- {d \over 2}]}
R_{i j} (\vec x, \vec y, \vec z)
\eeq
where $R_{ij}$ is the conformal tensor (\ref{JOOtensor}).

${ \bf \<J_i^a {\cal O}^I {  \cal O}^J\>}$: It is useful to
study the correlator 
$ \<J_i^a(\vec z) {\cal O}^I(\vec x)  {\cal O}^J (\vec y) \>$
from the $AdS$ viewpoint because the Ward 
identity~(\ref{ward}) which relates it to
$ \< {\cal O}(\vec y)^I  {\cal O}^J (\vec z) \>$
is a further check on the ${\rm CFT}/AdS$ conjecture. We assume that
${\cal O}^I(\vec x)$ 
is a scalar composite operator, in a real representation
of the $SO(6)$ flavor group with generators $\stackrel{a\ \,}{T^{IJ}}$
which are imaginary antisymmetric matrices, and that ${\cal O}^I(\vec x)$ 
corresponds to a real scalar field $\phi^I(\vec x)$ in
$AdS_5$ supergravity. Actually we will present an
$AdS_{d+1}$ calculation based on a gauge--invariant extension of
(\ref{scalaraction}), namely
\bea
\label{scalarcovariantaction}
S[\phi^I, A_{\mu}^a] &=&
\frac{1}{2} \, \int d^dz dz_0 \,\sqrt{g} \,
\left[
g^{\mu \nu} D_{\mu} \phi^I D_{\nu} \phi^I + m^2 \phi^I \phi^I 
\right]\\
D_{\mu} \phi^I & = & \partial_{\mu} \phi^I - i A_{\mu}^a 
\stackrel{a\ \,}{T^{IJ}} 
\phi^J \nonumber
\eea
The cubic vertex then leads to the $AdS$ integral representation
of the gauge theory correlator
\beq 
\label{JOO}
 \<J_i^a(\vec z) {\cal O}^I(\vec x)  {\cal O}^J (\vec y) \>
=  \stackrel{a\ \,}{T^{IJ}} 
\,\int \frac{d^d w dw_0}{w_0^{d+1}}
G_{\mu i}(w , \vec z) w_0^2
K_{\Delta}(w, \vec x) 
\frac{\stackrel{\leftrightarrow}{\partial}}{\partial w_{\mu}}
K_{\Delta}(w, \vec y)
\eeq
The integral is easily done by setting $\vec z = 0$ and applying
inversion. We have also shown that $\vec y= 0 $
followed by inversion gives the same final result,
which is 
\bea
\label{JOOfinal}
 \<J_i^a(\vec z) {\cal O}^I(\vec x)  {\cal O}^J (\vec y) \>
&=& \frac{2 C^d \stackrel{a\ \,}{T^{IJ}}}{|\vec x|^{2 \Delta}
|\vec y|^{2 \Delta}}
\frac{\partial}{\partial x'_i} \int \frac{d^d w' d w'_0}{w'_0}
K_{\Delta}(w', \vec{x'})K_{\Delta}(w', \vec{y'})
\\
& = & - \xi
\stackrel{a\ \,}{S^{IJ}}(\vec z, \vec x , \vec y) \nonumber
\\
\xi & = & \frac{(\Delta- \frac{d}{2}) \Gamma[{d \over 2}]
\Gamma[\Delta]}{\pi^d (d-2) \Gamma[\Delta - {d \over 2}]}
\nonumber
\eea
where $\stackrel{a\ \,}{S^{IJ}}(\vec z, \vec x , \vec y)$ 
is the conformal amplitude
of (\ref{JOOtensor}).
Comparing with (\ref{ward}) and (\ref{2pointsbyparts}), we see that the
expected Ward identity is {\it not} satisfied; there is a mismatch
by a factor $\frac{2 \Delta - d}{\Delta}$.
Although we have checked the integral thoroughly, this is an important point,
so we now give a heuristic argument that the answer is correct. We compute 
the divergence of the correlator (\ref{JOO}) using the following
identity inside the integral:
\beq
\frac{\partial}{\partial z_i} \, G_{\mu i}(w, \vec z) =
- \frac{\partial}{\partial w_{\mu}} \, K_d(w, \vec z)
\eeq
where $K_d(w, \vec z)$ is the Green's function of a massless scalar,
{\it i.e.} $\Delta=d$. If we integrate by parts, the bulk term vanishes
and we find
\beq
\frac{\partial}{\partial z_i}\,
\<J_i^a(\vec z) {\cal O}^I(\vec x)  {\cal O}^J (\vec y) \> 
 = 
\lim_{\epsilon \rightarrow 0} \,\int d^dw 
\epsilon^{1-d} K_{d}(\epsilon, \vec w, \vec z) 
\left[ K_{\Delta}(w , \vec x) 
\frac{\stackrel{\leftrightarrow}{\partial}}{\partial w_{0}}
K_{\Delta}(w , \vec y)
\right]_{w_0= \epsilon} 
\eeq
\beq
 =  - \left(
\frac{\Gamma[\Delta]}{ \pi^{d \over 2} \Gamma[\Delta - {d \over 2}]}
\right)^2 2 \Delta \lim_{\epsilon \rightarrow 0} \,\int d^dw 
\delta(\vec w - \vec z) 
\left[ \frac{ w_0^{2 \Delta-d+2}}{(w- \vec y)^{2(\Delta+1)}}
\frac{1}{(w - \vec x)^{2 \Delta}}
- (\vec x \leftrightarrow \vec y ) \right]_{w_0 = \epsilon} \nonumber
\eeq
where we used the property $\lim_{w_0 \rightarrow 0} K_d = \delta(\vec w
- \vec z) $ (see (\ref{rightsingular})). It also follows from
(\ref{scalarprop}--\ref{rightsingular}) that
\beq
\lim_{w_0 \rightarrow 0} 
\frac{w_0^{2 \Delta - d + 2}}{(w - \vec y)^{2(\Delta+1)}}
= \frac{ \pi^{d \over 2} \Gamma[ \Delta - \frac{d}{2} +1]}{\Gamma[\Delta +1]}
\, \delta^d(\vec w - \vec y)
\eeq
This gives 
\beq
\frac{\partial}{\partial z_i}\,
\<J_i^a(\vec z) {\cal O}^I(\vec x)  {\cal O}^J (\vec y) \> 
= \xi  \frac{(d-2) 2 \pi^{d \over 2} }{\Gamma[ {d \over 2}]}
 \stackrel{a\ \,}{T^{IJ}} 
\left(
\delta^d(\vec x - \vec z) - \delta^d(\vec y - \vec z)
\right)
\frac{1}{|\vec x - \vec y |^{2 \Delta}}
\eeq
which is consistent with (\ref{JOOfinal}) and confirms the previously 
found mismatch between $\<J_i^a {\cal O}^I  {\cal O}^J  \> $
and $\< {\cal O}^I  {\cal O}^J  \> $.

Thus the observed phenomenon is that the Ward identity relating
the correlators $\<J_i^a {\cal O}^I 
 {\cal O}^J  \> $ and
$\< {\cal O}^I  {\cal O}^J  \> $, as calculated from $AdS_{d+1}$
supergravity, is satisfied for  operators
${\cal O}^I $
of scale dimension $\Delta = d$, for which the corresponding $AdS_{d+1}$
scalar is massless, but fails for $\Delta \neq d$.

We suggest the following
interpretation of the problem, namely that
the prescription of \cite{witten} is correct for $n$--point correlators
in the boundary ${\rm CFT}_d$ for $n \ge 3$, but 2--point
 correlators are more singular, so a more careful procedure is required.
The fact that the kinetic and mass term integrals in (\ref{2pointscalars})
are each divergent has already been noted.
In the Appendix we outline an alternate calculation of 2--point functions,
very similar to that of \cite{polyakov}, in which we Fourier transform
in $\vec x$ and write a solution $\phi(z_0, \vec k)$ of the massive
scalar field equation which satisfies a Dirichlet boundary--value
problem at a small finite value $z_o = \epsilon$, compute the
2--point correlator at this value and then scale to $\epsilon =0$.
This procedure gives a value of $\< {\cal O}^I {\cal O}^J \> $
which is exactly a factor $\frac{2 \Delta -d}{\Delta}$ times
that of (\ref{2pointsbyparts}) and thus agrees with the Ward identity.

\def\ba{\begin{array}}
\def\ea{\end{array}}

\section*{Acknowledgements}
It is a pleasure to thank Edward Witten for prompt
reading of the manuscript and useful correspondence as well as
 Jeffrey Goldstone, Ken Johnson  
and Hermann Nicolai
for helpful discussions. 
The research of D.Z.F. is supported in part by
NSF Grant No. PHY-97-22072, S.D.M., A.M. and L.R. by D.O.E. cooperative agreement
DE-FC02-94ER40818. L.R. is  supported in part by INFN `Bruno Rossi' Fellowship.

\vspace*{.5in}

\section*{Appendix}
For scalars with dimension $\Delta=d$ the 
correlation functions 
achieve constant 
limiting values as we approach the boundary of $AdS$ space. If 
$\Delta \ne d$ then the correlation function goes 
to zero or infinity as we go towards the boundary, 
and must be defined with an appropriate scaling.
 In this case an interesting subtlety is seen 
to 
arise in the 
order in which we take the limits to define various quantities, 
and we discuss this issue below.

Let us discuss the 2--point function for scalars. We take the metric
 (\ref{lob}) on the $AdS$
 space, and put the boundary at $z_0=\epsilon$ with $\epsilon<<1$; 
at the end of the calculation we take $\epsilon$ to zero. We also 
Fourier transform the variables $\vec x$, and follow the discussion 
of \cite{polyakov}.

The wave equation in Fourier space for scalars with mass $m$ is
\be
z_0^{d+1}\frac{\partial}{\partial z_0}
[ z_0^{-d+1}\frac{\partial}{\partial z_0}\phi(z_0, \vec k)]-(k^2z_0^2+m^2)
\phi(z_0, \vec k)=0
\ee
where we have written
\be
\phi(z_0, \vec x)~=~{1\over (2\pi)^{d/2}}\int d\vec k ~e^{i
\vec{k}\cdot \vec{x}}~\phi(z_0, \vec k)
\ee

The solution to this equation is
\be
\phi(z_0, \vec k)~=~z_0^{d \over 2}F_\nu[ikz_0]
\ee
where $F_\nu$ is a solution of the Bessel equation with index 
\be
\nu~=~\Delta-{d\over  2} ~=~[{d^2\over 4}+m^2]^{1/2}
\ee

The action in terms of Fourier components is
\be
\ba{lll}
&&S=
{1\over 2}\int dz_0 \, 
d\vec k d \vec k'\delta(\vec k+\vec k')   z_0^{-d+1}\\ \\
&&[\frac{\partial}{\partial z_0}\phi(z_0,\vec k)\frac{\partial}{\partial z_0}
\phi(z_0,\vec k')+(k^2+{m^2\over z_0^2})\phi(z_0,\vec k)\phi(z_0,\vec k')]
\ea
\ee

We have to evaluate this action on a solution of the equation of motion with 
$\phi(\epsilon, \vec k) \equiv \phi_b(\vec k)$ given. 
An integration by parts gives
\be
S={1\over 2}\int d\vec k d \vec k'\delta(\vec k+\vec k') \lim_{z_0
\rightarrow \epsilon} 
z_0^{-d+1}[\phi(z_0,\vec k)\partial_{z_0}\phi(z_0,\vec k')]
\ee

If we have a solution to the wave equation
$K^\epsilon(z_0, \vec k)$ such that
\be
\lim_{z_0\rightarrow\epsilon}
K^\epsilon(z_0, \vec k)~=~1, ~~~\lim_{z_0\rightarrow\infty}K(z_0, \vec k)=0
\ee
then we can write the desired solution to the wave equation as
\be
\phi(z_0, \vec k)~=~K(z_0, \vec k)\phi_b(\vec k)
\ee

Then the 2-point function in Fourier space will be  given by
\be
\<{\cal O}(\vec k){\cal O}(\vec k')\>~=~-
\epsilon^{-d+1}\delta(\vec k+\vec k')\lim_{z_0\rightarrow 
\epsilon}\partial_{z_0} K^\epsilon(z_0, \vec k)
\label{correlator}
\ee

We have
\be
K^\epsilon(z_0, \vec k)~=~({z_0\over \epsilon})^{d/2}{{\cal K}_\nu(kz_0)
\over {\cal K}_\nu(k\epsilon) }
\label{propagator}
\ee
where ${\cal K}$ is the modified Bessel function which vanishes as $z_0\rightarrow\infty$.  For small argument ${\cal K}_\nu$ has the expansion
\be
{\cal K}_\nu(kz_0)~=~2^{\nu-1}\Gamma(\nu) ({kz_0})^{-\nu}[1+\dots]~-~2^{-\nu-1}{\Gamma(1-\nu)\over \nu}
({kz_0})^{\nu}[1+\dots]
\ee
where the terms represented by `$\dots$' are positive integer powers of $(kz_0)^2$. 
Then (\ref{correlator}) gives 
\be
\ba{lll}
&&\<{\cal O}(\vec k){\cal O}(\vec k')\>=\\ \\
&&~~~-\epsilon^{-d+1}\delta(\vec k+\vec k')\lim_{z_0\rightarrow\epsilon}
(\epsilon k)^{-d/2}\partial_{z_0}\displaystyle{{ ({
kz_0})^{-\nu+{d\over 2}}+\dots - 2^{-2\nu}{\Gamma(1-\nu)\over \Gamma(1+\nu)} 
({kz_0})^{\nu+{d\over 2}}+\dots\over
({k\epsilon})^{-\nu}+\dots -2^{-2\nu}{\Gamma(1-\nu)\over \Gamma(1+\nu)} 
({k\epsilon})^{\nu}+\dots  } }\\ \\
&&~~~=-\epsilon^{2(\Delta-d)}\delta(\vec k+\vec k')
k^{2\nu}2^{-2\nu}{\Gamma(1-\nu)\over \Gamma(1+\nu)}(2\nu) ~~+~~\dots
\label{result}
\ea
\ee
Here in the last line we have written 
only those terms that correspond to the
power law behavior of the correlator in 
position space, and further only the largest 
such terms in the limit $\epsilon\rightarrow 0$ 
have been kept. In particular we have dropped terms 
that are integer powers in $k^2$,  even though some of 
these terms are multiplied by a smaller power of $\epsilon$ than 
the term that we have kept. The reason for dropping these terms is that they  give 
delta--function 
contact terms in the correlator after transforming to position 
space, and we are interested here in the correlation function for separated points.

The result (\ref{result}) is the Fourier transform of the function
\be
{1\over \pi^{d/2}}\epsilon^{2(\Delta-d)}{(2\Delta-d)\over 
\Delta}{\Gamma(\Delta+1)\over \Gamma(\Delta-{d\over 2})}|\vec x-\vec y|^{-2\Delta}
\ee
which should therefore be the correctly normalized 
2--point function on the boundary $z_0=\epsilon$. It also 
agrees with the correctly normalized 
2--point function required by the Ward identity (\ref{ward}).
 The power of $\epsilon$ indicates the rate of growth of this 
correlation function as the boundary of $AdS$ space is moved to infinity, 
and we can define for convenience a scaled correlator that is the 
same as above but without this power of $\epsilon$. The 
correlation functions given in the rest of this paper are in fact written after such a
 rescaling.

We would however have obtained 
a different result had we taken the 
limits in the following way. 
We first take $\epsilon\rightarrow 0$ in the propagator 
(\ref{propagator}), obtaining
\be
K^\epsilon(z_0)~=~({z_0\over \epsilon})^{d/2}{1\over 
2^{\nu-1}\Gamma(\nu)(k\epsilon)^{-\nu}}{{\cal K}_\nu(kz_0)
}
\label{propagatorp}
\ee

Using (\ref{propagatorp}) in (\ref{correlator}) we get
\be
\<{\cal O}(\vec k){\cal O}(\vec k')\>=
-\epsilon^{-2(\Delta-d)}\delta(\vec k+\vec k')k^{2\nu}
2^{-2\nu}{\Gamma(1-\nu)\over \Gamma(1+\nu)}(\nu+{d\over 2})~~
+~~\dots
\label{resultp}
\ee
which differs from (\ref{result}) by a factor
\be
{\Delta\over 2\Delta-d}
\ee

The difference between (\ref{result}) and (\ref{resultp})
can be traced to the fact that the terms in $K^\epsilon(z_0)$ 
which are subleading in $\epsilon$ when $z_0$ is order unity, 
give a contribution that is not subleading when $z_0\rightarrow 
\epsilon$, which is the limit that we actually require when 
computing the 2-point function.


\newpage

\begin{thebibliography}{ll}
\bibitem{maldacena}J. Maldacena, `The Large $N$ Limit of Superconformal
Theories and Supergravity', hep--th/9711200
\bibitem{polyakov}{S.S. Gubser, I.R. Klebanov and A.M. Polyakov,
`Gauge Theory Correlators from Non--critical String Theory',
hep--th/9802109}
\bibitem{witten}{E. Witten, `Anti--de Sitter
Space and Holography', hep--th/9802150}
\bibitem{ferrarafronsdal}S. Ferrara, C. Fronsdal, `Gauge Fields as Composite
Boundary Excitations', hep-th/9802126.
\bibitem{nearhorizongeometry}{G. Gibbons, Nucl. Phys. B207 (1982) 337}
\bibitem{}{ R. Kallosh
and A. Peet, Phys. Rev. B46 (1992) 5223, hep-th/9209116}
\bibitem{}{ S. Ferrara,
G. Gibbons, R. Kallosh, Nucl. Phys.  B500 (1997) 75, hep-th/9702103.}
\bibitem{}{G. Gibbons and P. Townsend, `Vacuum Interpolation In Supergravity
Via Super $p$-Branes,' Phys. Rev. Lett. 71 (1993) 5223.}
\bibitem{}{M. J. Duff, G. W. Gibbons, and P. K. Townsend,
`Macroscopic Superstrings As Interpolating Solitons,'
Phys. Lett.  B332 (1994) 321.}
\bibitem{}{G. W. Gibbons, G. T. Horowitz, and P. K. Townsend,
`Higher Dimensional Resolution Of Dilatonic Black Hole Singularities,'
Class. Quant. Grav. 12 (1995) 297.}
\bibitem{}{S. Ferrara, G. W. Gibbons, and R. Kallosh, `Black Holes
And Critical Points In Moduli Space,' Nucl. Phys. B500 (1997) 75,
hep-th/9702103} 
\bibitem{nearhorizongeometryfinal}
{A. Chamseddine, S. Ferrara, G. W. Gibbons, and R. Kallosh,
`Enhancement Of Supersymmetry Near $5-D$ Black Hole Horizon,'
Phys. Rev. D55 (1997) 3647, hep-th/9610155.}


\bibitem{earlierappearance}{S. S. Gubser, I. R. Klebanov, and A. W. Peet,
`Entropy And Temperature Of Black 3-Branes,' Phys. Rev. D54
(1996) 3915.} 
\bibitem{}{I. R. Klebanov, `World Volume Approach To Absorption By
Nondilatonic Branes,' Nucl. Phys. B496 (1997) 231.}
\bibitem{}{S. S. Gubser, I. R. Klebanov, A. A. Tseytlin,
`String Theory And Classical Absorption By Three-branes,'
Nucl. Phys. B499 (1997) 217.}
\bibitem{}{S. S. Gubser and I. R. Klebanov, `Absorption By Branes And
Schwinger Terms In The World Volume Theory,' Phys. Lett. B413
(1997) 41.} 
\bibitem{earlierappearancefinal}{J. Maldacena and A. Strominger, 
 `Semiclassical Decay Of Near Extremal Fivebranes,'    hep-th/9710014.}



\bibitem{sugra}{H. J. Kim, L. J. Romans, and P. van Nieuwenhuizen,
`The Mass Spectrum Of Chiral $N=2$ $D=10$ Supergravity on $S^5$',
Phys. Rev. D32 (1985) 389.}
\bibitem{}{M. Gunaydin and N. Marcus, `The Spectrum Of The $S^5$
Compactification Of The Chiral $\N=2$ $D=10$ Supergravity And
The Unitary Supermultiplets Of $U(2,2|4)$,' Class. Quant. Grav.
 2 (1985) L11-17.}
\bibitem{sugrafinal}{M. Gunaydin, L.J. Romans, N.P. Warner,
`Compact and Non--Compact Gauged Supergravity Theories in Five
Dimensions', Nucl. Phys. B272, 598 (1986)}

\bibitem{bakerjohnson}{M. Baker and K. Johnson, Physica 96A, 120 (1979)}
\bibitem{anselmi}{D. Anselmi, D.Z. Freedman, M.T. Grisaru, A.A. Johansen,
`Universality of the operator product expansions of  $SCFT_4$',
Phys. Lett. B394 (1997) 329, hep--th/9608215}
\bibitem{anselmialone}{D. Anselmi, `Central Functions and Their
Physical Implications', hep-th/9702056}
\bibitem{positivity}{D. Anselmi, J. Erlich, D.Z. Freedman and A.A. Johansen,
`Positivity Constraints on Anomalies in Supersymmetric Gauge Theories', 
hep-th/9711035}
\bibitem{nonrenorm}{D. Anselmi, D.Z. Freedman, M.T. Grisaru, A.A. Johansen,
`Nonperturbative Formulas for Central Functions of Supersymmetric
Gauge Theories', hep-th/970842}
\bibitem{erlich}{J. Erlich and D.Z. Freedman, `Conformal Symmetry and the
Chiral Anomaly', Phys. Rev. D55 (1997) 5209--5217, hep-th/9611133}
\bibitem{osborn}{H. Osborn and A. Petkou, Ann. Phys.(N.Y) 231, 311
(1994)}
\bibitem{schreier}{E.J.Schreier, `Conformal Symmetry and the Three--point
Function', Phys. Rev. D 3, 980 (1971)}
\bibitem{dzfgrignani}{D.Z.Freedman, G.Grignani, N.Rius, K.Johnson,
`Conformal Symmetry and Differential Regularization of the Three--gluon
Vertex', Ann. Phys. 218 (1992) 75--120}
\bibitem{jackiw}{R. Jackiw, `Topological Investigations of Quantized
Gauge Theories', in `Current Algebra and Anomalies', eds.
S.B. Treiman, R. Jackiw, B. Zumino and E. Witten, Princeton University Press}
\bibitem{osbornprivate} H. Osborn, Private Communication
\bibitem{everybodyfirst}{N. Itzhaki, J. M. Maldacena, J. Sonnenschein, and
S. Yankielowicz, `Supergravity And The Large $N$ Limit Of Theories
With Sixteen Supercharges,' hep-th/9802042.}
\bibitem{}{S. Hyun, `$U$-Duality Between Three And Higher Dimensional
Black Holes,' hep-th/9704005}
\bibitem{}{K. Sfetsos and K. Skenderis, `Microscopic Derivation Of The
Bekenstein-Hawking Entropy Formula For Nonextremal Black Holes,'
hep-th/9711138}
\bibitem{}{ H. J. Boonstra, B. Peeters, and K. Skenderis,
`Branes And Anti-de Sitter Space-times,' hep-th/9801076.}
\bibitem{}{P. Claus, R. Kallosh, and 
A. van Proeyen, `$M$ Five-brane And Superconformal $(0,2)$ Tensor Multiplet
In Six-Dimensions,' hep-th/9711161} 
\bibitem{}{P. Claus, R. Kallosh, J. Kumar,
P. Townsend, and A. van Proeyen, `Conformal Theory Of $M2$, $D3$, $M5$,
and $D1$-Branes $+$ $D5$-Branes,' hep-th/9801206.}
\bibitem{}{R. Kallosh, J. Kumar, and A. Rajaraman, `Special Conformal
Symmetry Of Worldvolume Actions,' hep-th/9712073.}
\bibitem{}{S. Ferrara and C. Fronsdal, `Conformal Maxwell Theory As
A Singleton Field Theory On $AdS(5)$, IIB Three-Branes And Duality,'
hep-th/9712239.}
\bibitem{}{M. Gunaydin and D. Minic, `Singletons, Doubletons and $M$
Theory,' hep-th/9802047.}
\bibitem{}{S.P. de Alwis,
`Supergravity The DBI Action And Black Hole Physics',
hep-th/9804019}
\bibitem{}{S. Ferrara, A. Kehagias, H. Partouche, A. Zaffaroni,
`Ads(6) Interpretation Of 5-D Superconformal Field Theories', hep-th/9804006}  
\bibitem{}{Andreas Brandhuber, Nissan Itzhaki, Jacob Sonnenschein, 
Shimon Yankielowicz,
`Wilson Loops, Confinement, and Phase Transitions In Large N Gauge
Theories From Supergravity', hep-th/9803263 }
\bibitem{}{Miao Li, `'t Hooft Vortices On D-Branes', hep-th/9803252}
\bibitem{sfetsos}{Mans Henningson, Konstadinos Sfetsos, 
`Spinors And The AdS / CFT Correspondence', hep-th/9803251}
\bibitem{}{Harm Jan Boonstra, Bas Peeters, Kostas Skenderis,
`Brane Intersections, Anti-De Sitter Space-Times And Dual
Superconformal Theories',  hep-th/9803231}
\bibitem{}{Anastasia Volovich, 
`Near Anti-De Sitter Geometry And Corrections To The Large N
Wilson Loop', hep-th/9803220}
\bibitem{}{Zurab Kakushadze,
`Gauge Theories From Orientifolds And Large N Limit', hep-th/9803214}
\bibitem{}{L. Andrianopoli, S. Ferrara
`K-K Excitations On Ads(5)$\times S^5$ as N=4 'primary' Superfields',
hep-th/9803171}
\bibitem{}{Yaron Oz, John Terning,
`Orbifolds Of Ads(5)$\times S^5$ and 4-D Conformal Field Theories',
hep-th/9803167}
\bibitem{}{M. Gunaydin,
`Unitary Supermultiplets Of Osp(1/32,R) And M Theory', hep-th/9803138}
\bibitem{}{A. Brandhuber, N. Itzhaki, J. Sonnenschein, S. Yankielowicz,
`Wilson Loops In The Large N Limit At Finite Temperature', hep-th/9803137}
\bibitem{}{Soo-Jong Rey, Stefan Theisen, Jung-Tay Yee,
`Wilson-Polyakov Loop At Finite Temperature In Large N Gauge
Theory And Anti-De Sitter Supergravity', hep-th/9803135}
\bibitem{}{Edward Witten,
`Anti-De Sitter Space, Thermal Phase Transition, And Confinement
In Gauge Theories', hep-th/9803131}
\bibitem{}{Jaume Gomis, `Anti-De Sitter Geometry And Strongly Coupled 
Gauge Theories', hep-th/9803119}
\bibitem{}{Joseph A. Minahan,
`Quark - Monopole Potentials In Large N SuperYang-Mills', hep-th/9803111}
\bibitem{}{S. Ferrara, A. Kehagias, H. Partouche, A. Zaffaroni,
`Membranes And Five-Branes With Lower Supersymmetry And Their
AdS Supergravity Duals', hep-th/9803109}
\bibitem{}{E. Bergshoeff, K. Behrndt,
`D - Instantons And Asymptotic Geometries', hep-th/9803090}
\bibitem{}{Gary T. Horowitz, Simon F. Ross,
`Possible Resolution Of Black Hole Singularities From Large N
Gauge Theory', hep-th/9803085}
\bibitem{}{Edi Halyo,
`Supergravity On AdS$(4/7)\times S^{7/4}$ and M Branes', hep-th/9803077}
\bibitem{}{Michael Bershadsky, Zurab Kakushadze, Cumrun Vafa,
`String Expansion As Large N Expansion Of Gauge Theories', hep-th/9803076}
\bibitem{}{Robert G. Leigh, Moshe Rozali,
`The Large N Limit Of The (2,0) Superconformal Field Theory', hep-th/9803068}
\bibitem{}{M.J. Duff, H. Lu, C.N. Pope,
`$AdS_5\times S^5$ Untwisted', hep-th/9803061}
\bibitem{}{Sergio Ferrara, Alberto Zaffaroni,
`N=1, N=2 4-D Superconformal Field Theories And Supergravity In
AdS(5)', hep-th/9803060}
\bibitem{}{Shiraz Minwalla,
`Particles On Ads(4/7) And Primary Operators On M(2)-Brane And
M(5)-Brane World Volumes',  hep-th/9803053 }
\bibitem{}{Ofer Aharony, Yaron Oz, Zheng Yin,
`M Theory On $AdS_p\times S^{11-p}$ and Superconformal Field Theories',
hep-th/9803051}
\bibitem{}{Leonardo Castellani, Anna Ceresole, Riccardo D'Auria, Sergio 
Ferrara, Pietro Fre, Mario Trigiante,
`G/H M-Branes And Ads(P+2) Geometries', hep-th/9803039 }
\bibitem{}{I.Ya. Aref'eva, I.V. Volovich,
`On Large N Conformal Theories, Field Theories In Anti-De Sitter
Space And Singletons', hep-th/9803028}
\bibitem{}{Steven S. Gubser, Akikazu Hashimoto, Igor R. Klebanov,
 Michael Krasnitz,
`Scalar Absorption And The Breaking Of The World Volume
Conformal Invariance', hep-th/9803023 }
\bibitem{}{Moshe Flato, Christian Fronsdal,
`Interacting Singletons', hep-th/9803013 }
\bibitem{}{Juan Maldacena,
`Wilson Loops In Large N Field Theories', hep-th/9803002}
\bibitem{}{Soo-Jong Rey, Jungtay Yee,
`Macroscopic Strings As Heavy Quarks In Large N Gauge Theory And
Anti-De Sitter Supergravity', hep-th/9803001}
\bibitem{}{Sergio Ferrara, Christian Fronsdal, Alberto Zaffaroni,
`On N=8 Supergravity On Ads(5) And N=4 Superconformal Yang-Mills
Theory', hep-th/9802203 }
\bibitem{}{Micha Berkooz,
`A Supergravity Dual Of A (1,0) Field Theory In Six-Dimensions', 
hep-th/9802195}
\bibitem{everybodylast}{Shamit Kachru, Eva Silverstein,
`4-D Conformal Theories And Strings On Orbifolds', hep-th/9802183}
\bibitem{canadians}W. Muck, K. Viswanathan, `Conformal Field Theory
Correlators from Classical Scalar Field Theory on $AdS_{d+1}$',
hep--th/9804035.
\end{thebibliography}
\end{document}